\documentclass[a4paper, 11pt]{article}

\usepackage{jcappub}
\usepackage[top=30truemm,bottom=40truemm,left=20truemm,right=20truemm]{geometry}

\usepackage[T1]{fontenc}
\usepackage{newpxtext, newpxmath}
\usepackage{graphicx}

\allowdisplaybreaks[4]

\title{\bf Curvaton distribution from stochastic inflation}
\author{Koki Tokeshi}
\affiliation{\it Institute for Cosmic Ray Research (ICRR), The University of Tokyo, \\ 5-1-5 Kashiwanoha, Kashiwa, Chiba 277-8582, Japan}
\emailAdd{tokeshi@icrr.u-tokyo.ac.jp}

\abstract{
	The curvaton paradigm can realise a part of or all the observed curvature perturbation.  
	Based on the stochastic formalism of inflation and closed-form exact distributions therein, the distribution of the curvature perturbation is presented in an analytical manner by identifying a test field with a curvaton. 
	The parameter space consisting of the decay rate and the mass of the curvaton is studied to discuss the probability that the curvaton can contribute to the curvature perturbation in a non-negligible way. 
}

\begin{document}

\maketitle
\flushbottom

\section{Introduction}
\label{sec:intro}

Cosmic inflation~\cite{Starobinsky:1980te, Sato:1980yn, Guth:1980zm, Linde:1981mu, Linde:1983gd, Albrecht:1982wi} is now the leading paradigm that describes the earliest history of our universe. 
It is an era of a quasi-de-Sitter stage and dynamically avoids the fine-tuning issues of the standard hot Big Bang theory, such as the horizon, flatness, and unwanted-relics problems. 
Vacuum quantum fluctuations generated during inflation were amplified due to gravitational instability, giving rise to the seed of all the present cosmological structures. 
The nearly scale-invariant power spectrum of the curvature perturbation theoretically predicted has been confirmed to a high accuracy by various large-scale observations~\cite{Planck:2018vyg, Planck:2018jri, Planck:2019kim}. 

While in most cosmological scenarios the curvature perturbation originates from the classicalised fluctuations of a single scalar field, the so-called inflaton field, the presence of additional fields in the early universe is motivated by high-energy model constructions~\cite{Baumann:2014nda}. 
Amongst such scenarios, the curvaton paradigm~\cite{Lyth:2001nq, Moroi:2001ct, Enqvist:2001zp, Lyth:2002my} may also explain a part of or all the observed amount of curvature perturbation. 
This is because the isocurvature fluctuation of a curvaton is transformed into the adiabatic curvature perturbation after inflation ends. 
This implies that the inflaton field does not have to be responsible for generating it if the curvaton field generates a sufficient amount of the observed curvature perturbation, resulting in looser constraints on the inflationary models. 
Curvaton models can also generate a sizable amount of non-Gaussianity~\cite{Lyth:2002my, PhysRevD.69.043503, Lyth:2005fi, PhysRevD.71.123508, Enqvist:2005pg, Malik:2006pm, Sasaki:2006kq, Beltran:2008aa, Moroi:2008nn, PhysRevD.78.063545, Li:2008jn, Huang:2008ze, Huang:2008zj, Huang:2008bg, Huang:2008qf, Enqvist:2008gk, Kawasaki:2008pa, Chingangbam:2009xi, Chambers:2009ki, Enqvist:2009zf, Enqvist:2009ww, Byrnes:2010xd, Enqvist:2010ky, Byrnes:2011gh, Kawasaki:2011pd}, which is one of the detectable signals in the observables. 
Eccentric objects such as primordial black holes~\cite{Zeldovich:1967lct, Hawking:1971ei, Carr:1974nx, Carr:1975qj} could also be generated in curvaton scenarios~\cite{Kohri:2012yw, Kawasaki:2013xsa, Bugaev:2013vba, Young:2013oia, PhysRevD.87.063519, Ando:2017veq, Ando:2018nge, Chen:2019zza, PhysRevLett.126.131301, Kawasaki:2021ycf, Pi:2021dft, Liu:2021rgq, Meng:2022low, Ferrante:2023bgz, Inomata:2023drn, Gow:2023zzp} to explain the recently observed spectrum of gravitational waves. 

The conversion from the curvaton to the curvature perturbation enables us to have a functional relation between them. 
In other words, the information of the curvaton during inflation is handed down to the curvature perturbation. 
When the stochastic formalism of inflation~\cite{Starobinsky:1986fx} is applied in particular, it enables us to derive the distribution function of the curvature perturbation from the distribution function of the curvaton during inflation. 
Meanwhile, a class of all the possible exact solutions to a test (or spectator) field, such as a curvaton, in stochastic inflation has been presented in Ref.~\cite{Honda:2024evc} by virtue of the correspondence between the Fokker--Planck and imaginary-time Schr\"{o}dinger equations, together with isospectral Hamiltonians and an underlying symmetry called shape invariance~\cite{Cooper:1994eh, 10.1093/qmath/6.1.121, Gendenshtein:1983skv}. 
The list contains of course the simplest Gaussian case (that is, a test field in the quadratic potential), as well as three other cases in which the corresponding quantum-mechanical systems can be solved exactly in terms of classical orthogonal polynomials. 
Each of the exact statistical quantities such as the distribution and correlation functions enables us to study the test-field dynamics and phenomenology both during and after inflation in an analytical manner.

This paper thus aims to present one of the possible applications of those exact solutions, focussing on a curvaton scenario. 
By identifying the test field during inflation with the curvaton, the analytical expressions for the distribution function of the curvature perturbation will be derived. 
While the same procedure applies even when the time evolution of the distribution function is taken into account, this paper restricts itself to the stationary state, where the initial condition is washed away, in order to give a simpler demonstration of the exact solutions. 
Since the amount of the generated curvature perturbation depends on the energy fraction of the curvaton, there can be two situations where the curvaton dominates or does not dominate the universe before its decay. 
The distribution correspondingly includes the two branches that come from those two situations. 
The interplay amongst the model parameters, the decay rate and mass of the curvaton, and the curvature perturbation gives rise to the non-trivial structure of the parameter space, which will also be studied. 
It is found that there exist regions to which a non-negligible contribution from the curvaton to the curvature perturbation can be realised. 
The way of analyses presented in the present paper basically follows~\cite{Enqvist:2012xn, Lerner_2014} (see also Refs.~\cite{Lyth:2006gd, Lyth:2007jh}), in which deriving the distribution of the curvature perturbation from that of the curvaton in stochastic inflation was first performed. 
The parameter-space structure was in particular analysed in~\cite{Lerner_2014}, focussing on the curvaton in the simplest quadratic potential and given that the observed amount of the curvature perturbation is of order $10^{-5}$. 
While in the references the non-Gaussianity parametrised by $f_{\rm NL}$ was taken into account, it will not be considered in the present paper to make our first demonstration of the exact solutions as simple as possible. 

The present paper is organised as follows. 
The basics of the stochastic formalism of inflation as well as a class of the possible exact solutions therein, mainly focussing on the stationary (equilibrium) solutions, are reviewed in section~\ref{sec:stochinf}. 
Application of them to a curvaton scenario is presented in section~\ref{sec:curvaton}, where the probability distribution function of the curvature perturbation is derived analytically. 
The parameter-space structure in light of the observed amount of the curvature perturbation will also be studied in section~\ref{sec:curvaton}. 
Section~\ref{sec:concl} is devoted to summarising the result. 
Natural units ($c = \hbar = 1$) are used throughout, and $M_{\rm P} = 1 / \sqrt{8 \pi G} \simeq 2.4 \times 10^{18} \, \mathrm{GeV}$ denotes the reduced Planck mass. 

\section{Stochastic inflation and exact solutions}
\label{sec:stochinf}

The standard cosmological perturbation theory assumes that the quantised fluctuation $\delta \phi (t, \, \vb*{x})$ is perturbatively small compared to the classical and homogeneous background $\overline{\phi} (t)$. 
Those quantum fluctuations are stretched out and continuously cross the horizon to stochastically affect the evolution of large-scale fields, giving rise to the so-called stochastic formalism of inflation~\cite{Starobinsky:1986fx}. 
It is based on gradient expansion~\cite{Salopek:1990jq, Deruelle:1994iz} rather than perturbative expansion with respect to the fields, and thus enables us to describe the evolution of the large-scale fields in a non-perturbative way. 
The large-scale primordial fluctuations can be computed by the (stochastic) $\delta \mathcal{N}$ formalism~\cite{Starobinsky:1985ibc, Sasaki:1995aw, Sasaki:1998ug, Lyth:2004gb, Fujita:2013cna, Fujita:2014tja}. 

\subsection{Test-field dynamics during inflation}
\label{sec:testfield}

Throughout this paper, it is assumed that the inflaton field maintains the nearly de-Sitter behaviour of the expanding universe, while the dynamics of a test (spectator) field $\phi$, different from the inflaton, is focussed on. 
It is also assumed that the test field interacts with gravity only minimally, but the variation of the Hubble parameter in time is negligible for the nearly massless test field. 

On a homogeneous and isotropic background, the standard Klein--Gordon equation describes the motion of $\phi$, 
\begin{equation}
    \dv[2]{\phi}{N} + 3 \dv{\phi}{N} - \frac{ \grad^{2} \phi}{ (a H)^{2} } + \frac{1}{H^{2}} \dv{V}{\phi} = 0 
    \,\, . 
    \label{eq:stochinf_eom} 
\end{equation}
In Eq.~(\ref{eq:stochinf_eom}), $a$ and $H = \dd \ln a / \dd t$ denote the scale factor of the space-time and the Hubble parameter, respectively, and $V$ is the potential. 
Here and hereafter, $N$, the number of $e$-folds, is used instead of the cosmic time $t$~\cite{Vennin:2015hra}. 
Introducing the time-dependent cut-off by $k_{\sigma} (N) \coloneqq \sigma \cdot a H$ ($\sigma$ is a constant that satisfies $0 < \sigma \ll 1$), the effective equation of motion for the large-scale field configurations is derived through the scale decomposition of the field, 
\begin{equation}
    \phi (N, \, \vb*{x})  
    = \phi_{<} (N, \, \vb*{x}) + \phi_{>} (N, \, \vb*{x})
    \,\, , 
    \label{eq:stochinf_decomp}
\end{equation}
where each component defines the classicalised large-scale modes and the sub-horizon fluctuations,
\begin{subequations}
    \label{eq:stochinf_decomp_def}
    \begin{align}
        \phi_{<} (N, \, \vb*{x})
        &\coloneqq 
        \int \frac{\dd^{3} k}{ (2 \pi)^{3} } \qty{ 1 - \widetilde{\mathcal{W}} \qty[ \frac{k}{k_{\sigma} (N)} ] } \widetilde{\phi} (N, \, \vb*{k}) e^{i \vb*{k} \cdot \vb*{x}} 
        \,\, , 
        \label{eq:stochinf_decomp_defa}
        \\ 
        \phi_{>} (N, \, \vb*{x}) 
        &\coloneqq  
        \int \frac{\dd^{3} k}{ (2 \pi)^{3} } \widetilde{\mathcal{W}} \qty[ \frac{k}{k_{\sigma} (N)} ] \widetilde{\phi} (N, \, \vb*{k}) e^{i \vb*{k} \cdot \vb*{x}} 
        \,\, . 
        \label{eq:stochinf_decomp_defb}
    \end{align}
\end{subequations}
The window function $\widetilde{\mathcal{W}}$ has been inserted in Eqs.~(\ref{eq:stochinf_decomp_def}). 
Of several choices for the window function, the Heaviside step function $\widetilde{\mathcal{W}} (z) = \Theta (z - 1)$ has widely been used for its simplicity, and is therefore applied here. 
This specific implementation enables us to arrive at the Langevin equation in the slow-roll regime, 
\begin{equation}
    \dv{\phi}{N} 
    = - \frac{1}{3 H^{2}} \dv{V}{\phi} + \frac{H}{2 \pi} \xi
    \,\, . 
    \label{eq:stochinf_eom_lan}
\end{equation}
In Eq.~(\ref{eq:stochinf_eom_lan}) and hereafter, the large-scale part of the field defined in Eq.~(\ref{eq:stochinf_decomp_defa}) is simply denoted by $\phi$. 
The Langevin equation is one of the fundamental equations in stochastic inflation, which reduces to the conventional and deterministic slow-roll equation if one neglects the second term in the right-hand side. 
The second term is a random noise with the amplitude $H / 2 \pi$, with the statistical properties $\expval{ \xi (N, \, \vb*{x}) } = 0$ and $\expval{ \xi (N_{1}, \, \vb*{x}_{1}) \xi (N_{2}, \, \vb*{x}_{2}) } = \mathrm{sinc} \qty( k_{\sigma} \abs{ \vb*{x}_{1} - \vb*{x}_{2} } ) \, \delta_{\rm D} (N_{1} - N_{2})$. 
The appearance of the Dirac $\delta$-function comes from the specific choice of the window function (see Refs.~\cite{Hu:1992ig, Casini:1998wr, Winitzki:1999ve, Matarrese:2003ye, Liguori:2004fa, Breuer:2006cd, Mahbub:2022osb} for other choices), since the derivative with respect to time hits the window function. 
This means that at each time only the mode with wavenumber $k_{\sigma}$ crosses the horizon to become a part of the large-scale modes. 
Each random realisation generated by Eq.~(\ref{eq:stochinf_eom_lan}) describes the evolution of the field in each Hubble region at the leading order of gradient expansion (so each Hubble patch evolves ``separately''). 
This is why in the following the limit $\vb*{x}_{1} = \vb*{x}_{2}$ (one-point statistics) will only be of interest. 

The Langevin equation~(\ref{eq:stochinf_eom_lan}) gives rise to the governing equation for the distribution function of the stochastic field $\phi$. 
It is nothing but the Fokker--Planck equation,~\cite{risken1989fpe}
\begin{equation}
    \pdv{f}{N} 
    = \qty( \frac{1}{3 H^{2}} \pdv{\phi} \dv{V}{\phi} + \frac{H^{2}}{8 \pi^{2}} \pdv[2]{\phi} ) f 
    \eqqcolon \mathcal{L}_{\rm FP} f 
    \,\, . 
    \label{eq:stochinf_fp}
\end{equation}
Here, $f = f (\phi, \, N) = f (\phi, \, N \mid \phi_{0}, \, N_{0})$ is the transition probability that tells the distribution of $\phi$ at a given time $N$, starting from an initial condition $(\phi_{0}, \, N_{0})$. 
Since Eq.~(\ref{eq:stochinf_fp}) is a partial differential equation linear in $f$, the solution has the form of spectral expansion, 
\begin{equation}
    f (\phi, \, N) 
    = \sum_{n = 0}^{\infty} c_{n} \exp \qty[ - \frac{4 \pi^{2}}{3 H^{4}} V (\phi) ] \Psi_{n} (\phi) \exp \qty[- \lambda_{n} (N - N_{0})]  
    \,\, .
    \label{eq:stochinf_specexpa}
\end{equation}
The expansion coefficient $c_{n}$ is determined by imposing an initial condition, assumed to be concentrated, \textit{i.e.}~$f (\phi, \, N_{0}) = \delta_{\rm D} (\phi - \phi_{0})$. 
The decay rate in the exponent $\lambda_{n}$ is the $n$-th eigenvalue of the Fokker--Planck operator $\mathcal{L}_{\rm FP}$. 
The absence of the continuous spectrum has been assumed in Eq.~(\ref{eq:stochinf_specexpa}). 
In other words, though there are both discrete and continuous eigenstates in general, situations in which only the discrete states exist will be focussed on to simplify our analysis in the following, as was done in Ref.~\cite{Honda:2024evc}. 

The second-order ordinary differential equation for $\Psi_{n} (\phi)$ follows from Eq.~(\ref{eq:stochinf_fp}) that 
\begin{equation}
    \qty[ - \frac{ H^{2} }{ 8 \pi^{2} } \dv[2]{\phi} + V_{\rm S} (\phi) ] \Psi_{n} (\phi) = \lambda_{n} \Psi_{n} (\phi) 
    \,\, . 
    \label{eq:stochinf_sch}
\end{equation}
This is the same form as the stationary Schr\"{o}dinger equation in non-relativistic quantum mechanics, where the positive-semidefiniteness Hamiltonian $\mathsf{H} \coloneqq - (H^{2} / 8 \pi^{2}) \, \dd^{2} / \dd \phi^{2} + V_{\rm S} (\phi)$ is assumed. 
The Schr\"{o}dinger potential is related to the potential for the test field by  
\begin{equation}
    V_{\rm S} 
    \coloneqq \frac{2 \pi^{2}}{9 H^{6}} \qty( \dv{V}{\phi} )^{2} - \frac{1}{6 H^{2}} \dv[2]{V}{\phi} 
    \,\, . 
    \label{eq:stochinf_schpot}
\end{equation}
The domain of $\phi$, denoted by $\phi \in (\phi_{1}, \, \phi_{2})$, in general depends on the potential.
When $\phi$ is bounded in the sense that the potential satisfies $V_{\rm S} \to + \infty$ at both boundaries, there exist only discrete-energy states. 
When, on the other hand, $V_{\rm S}$ goes to some finite value as $\phi$ approaches a boundary, $\lambda_{n}$ can either be discrete or continuous. 
As mentioned below Eq.~(\ref{eq:stochinf_specexpa}), only the former cases will be considered throughout this paper. 

In the late-time limit, $N \to \infty$, the distribution function asymptotes to the stationary solution (as known as the Starobinsky--Yokoyama equilibrium formula~\cite{Starobinsky:1994bd}) satisfying $\partial f / \partial N = 0$, 
\begin{equation}
    f_{\infty} (\phi) 
    \coloneqq \lim_{N \to \infty} f (\phi, \, N) 
    = C \exp \qty[ - \frac{ 8 \pi^{2} }{ 3 H^{4} } V (\phi) ] 
    \,\, , 
    \qquad 
    \frac{1}{C} \coloneqq \int \dd \phi \, \exp \qty[ - \frac{ 8 \pi^{2} }{ 3 H^{4} } V (\phi) ] 
    \,\, . 
    \label{eq:stochinf_statdist}
\end{equation}
This stationary solution corresponds to the normalised ground-state wavefunction squared to Eq.~(\ref{eq:stochinf_sch}). 

\subsection{Exact solutions in stochastic inflation}
\label{subsec:exactsol}

The fact that there is the correspondence between the Fokker--Planck equation in stochastic inflation and Schr\"{o}dinger equation in non-relativistic quantum mechanics implies that all the possible and analytical expressions for the statistical quantities such as the distribution and correlation functions in stochastic inflation can exhaustively be constructed, as was done in Ref.~\cite{Honda:2024evc}. 
For practical use, four exact solutions out of the ten were constructed in Ref.~\cite{Honda:2024evc} by means of isospectral Hamiltonians and an underlying symmetry called shape invariance. 
Here, those solutions, focussing on the stationary solutions in particular, are reviewed with the application to a curvaton scenario in section~\ref{sec:curvaton} in mind. 

\begin{table}
	\centering 
	\renewcommand{\arraystretch}{2.0}
	\caption{
		The four exactly solvable situations for a test field in stochastic inflation. 
        The corresponding quantum-mechanical system can be solved exactly in terms of either the Hermite (H), Laguerre (L), or Jacobi (J) polynomial. 
		\\ 
	}
	\begin{tabular}{|c|c|c|c|}
		\hline 
		\textbf{Name} & \textbf{Class} & \textbf{Test-Field Potential} $v (y)$ & \textbf{Wavefunction} $\Psi_{n} (\phi)$ \\ 
		\hline \hline 
		HO & H & $y^{2} / 2$ & $e^{-z^{2} / 2} H_{n} (z)$ \\ 
		\hline 
		RHO & L & $y^{2} / 2 - (\ell + 1) \ln y$ & $z^{\qty( \ell + 1 )/2} e^{-z / 2} L_n^{\ell + 1/2} (z)$ \\ 
		\hline 
		Sc-I & J & $- s \ln \cos y + t \ln \qty[ \frac{\displaystyle 1 - \tan (y/2) }{ \displaystyle 1 + \tan (y/2) } ]$ & $\begin{array} {lcl} &{}& (1 - z)^{(s-t)/2} (1 + z)^{(s+t)/2} \\ &\quad& \times P_{n}^{s - t - 1/2, \, s + t - 1/2} (z) \end{array}$ \\ 
		\hline 
		RMI & J & $\displaystyle - \frac{t}{s} y - s \ln \cos y$ & $\begin{array} {lcl} &{}& \displaystyle \exp \qty( \frac{t}{n+s} \mathrm{arctan} z ) (1+z^{2})^{-(n+s)/2} \\ &\quad& \times P_{n}^{- (n+s) - i t / (n+s), \, - (n+s) + i t / (n+s)} (- i z) \end{array}$ \\ 
		\hline 
	\end{tabular}
	\label{tab:exsfour}
\end{table}

For an exactly solvable quantum-mechanical system, the potential of the test field during inflation in general takes the form of 
\begin{equation}
    \frac{V (\phi)}{H^4} 
    = \frac{3}{4 \pi^{2}} \qty[ v (y) - v (\widetilde{y}) ]
    \,\, , 
    \qquad 
    y \coloneqq \alpha \sqrt{ \frac{ 8 \pi^{2} }{ H^{2} } } \, \phi 
    \,\, . 
    \label{eq:exs_specv}
\end{equation}
It will be seen later that $\alpha$ is (can be) identified with the mass of the test field. 
The concrete form of the function $v (y)$ depends on the situation (see Table~\ref{tab:exsfour}), but in any case $\widetilde{y}$ is the location at which $v (y)$ becomes its global minimum, and the potential is assumed to vanish there. 
There are four exactly solvable situations where only the discrete states appear in the spectral decomposition~(\ref{eq:stochinf_specexpa}), summarised in Table~\ref{tab:exsfour}. 
Each of them is reviewed from now on. 

\paragraph{\textit{Harmonic oscillator}} 
The first exactly solvable situation consists of a test field confined in the quadratic potential, $v (y) = y^{2} / 2$. 
In this case, the corresponding Schr\"{o}dinger potential also takes the quadratic form. 
The normalised wavefunction, denoted by $\widehat{\Psi}_{n} (\phi)$ hereafter, is thus given by the Hermite polynomial, 
\begin{equation}
	\widehat{\Psi}_{n} (\phi) 
	= \frac{1}{\sqrt{2^{n} n! \sqrt{\pi}}} \, \qty( \alpha \sqrt{ \frac{8 \pi^{2}}{H^{2}} } \, )^{1/2} e^{- z^{2} / 2} H_{n} (z) 
	\,\, . 
	\label{eq:exs_wf_ho}
\end{equation}
In Eq.~(\ref{eq:exs_wf_ho}), $z \coloneqq y$ is called the sinusoidal coordinate. 
Substituting Eq.~(\ref{eq:exs_wf_ho}) into Eq.~(\ref{eq:stochinf_specexpa}) and employing the initial condition enables us to extract the coefficient $c_{n}$. 
The infinite summation can be reduced to the closed form by virtue of Mehler's summation formula for the Hermite polynomials~\cite{gradshteyn2007}, giving rise to 
\begin{equation}
	f (\phi, \, N) 
	= \frac{1}{\sqrt{\pi}} \qty( \alpha \sqrt{ \frac{8 \pi^{2}}{H^{2}} } \, )
    	e^{- z^2} 
    	\frac{\displaystyle \exp \qty{ \frac{1}{\sinh \qty[ 2 \alpha^{2} \qty( N - N_0 ) ]} \qty[ z z_0 - \frac{e^{- 2 \alpha^{2} \qty(N - N_0)}}{2} \qty( z^2 + z_0^2 ) ] } 
    	}{\sqrt{1 - e^{- 4 \alpha^{2} \qty(N - N_0)}}} 
	\,\, . 
	\label{eq:exs_dist_ho}
\end{equation}
The expression (\ref{eq:exs_dist_ho}) admits to a Gaussian distribution by completing the square, with the mean and variance given by 
\begin{subequations}
	\label{eq:exs_mv}
	\begin{align}
		\expval{ \phi } (N) 
		&= \phi_{0} e^{- 2 \alpha^{2} (N - N_{0})} 
		\,\, , 
        \label{eq:exs_mv_a}
        \\ 
		\qty[ \expval{\phi^{2}} - \expval{\phi}^{2} ] (N) 
		&= \frac{1}{4 \alpha^{2}} \qty( \frac{H}{2 \pi} )^{2} \qty[ 1 - e^{- 4 \alpha^{2} (N - N_{0}) } ] 
		\,\, . 
        \label{eq:exs_mv_b}
	\end{align}
\end{subequations}
In the late time limit, $N \to \infty$, the higher modes decay and only the zeromode survives, 
\begin{equation}
	f_{\infty} (\phi) 
    	= \qty[ \widehat{\Psi}_{0} (\phi) ]^{2}
    	= \frac{1}{\sqrt{\pi}} \qty( \alpha \sqrt{ \frac{8 \pi^{2}}{H^{2}} } \, ) e^{- z^{2}} 
    	\,\, . 
	\label{eq:exs_stdist_ho}
\end{equation}
When the exact solutions are applied to a curvaton scenario in section~\ref{sec:curvaton}, it is assumed that the inflationary stage has already settled down to the stationary state described by Eq.~(\ref{eq:exs_stdist_ho}) in the present case. 
In other words, the total duration of inflation (the total number of $e$-folds) is assumed to be much longer than $N_{\star} = 60$ (see, however, Refs.~\cite{Enqvist:2012xn, Hardwick:2017fjo}), when the distribution of the curvaton is derived. 

\paragraph{\textit{Radial harmonic oscillator}}
A test field confined in the potential $v (y) = y^{2} / 2 - (\ell + 1) \ln y$ is also exactly solvable, which corresponds to the quantum-mechanical system of the radial component of the three-dimensional harmonic oscillator. 
The wavefunction is thus given in terms of the associated Laguerre polynomial,\footnote{
    In $\ell \to -1$ limit, the associated Laguerre polynomial and the Hermite polynomial are related through the relation~\cite{abramowitz+stegun} 
    \begin{equation}
        H_{2 n} (z)  
        = (-1)^{n} 2^{2 n} n! L_{n}^{-1/2} (z^{2}) 
        \,\, . 
    \end{equation}
    This implies that in the limit only the even-mode states of the pure harmonic oscillator are reproduced from the radial harmonic oscillator. 
} 
\begin{equation}
    \widehat{\Psi}_{n} (\phi) 
    = \qty( \sqrt{ \frac{\omega}{2} }  \sqrt{ \frac{ 8 \pi^{2} }{ H^{2} } } \, )^{1/2} 
    \sqrt{ \frac{ 2 n! }{ \Gamma (n + \ell + 3/2) } } \, z^{(\ell + 1)/2} e^{- z/2} L_{n}^{\ell + 1/2} (z) 
    \,\, . 
    \label{eq:exs_rho_wf}
\end{equation} 
The Hardy--Hille summation formula~\cite{gradshteyn2007} enables us to let the infinite summation (\ref{eq:stochinf_specexpa}) be closed form, 
\begin{align}
    f (\phi, \, N) 
    &= 2 \qty( \sqrt{ \frac{\omega}{2} }  \sqrt{ \frac{ 8 \pi^{2} }{ H^{2} } } \, ) 
    z^{\ell + 1} e^{- z} 
    \frac{
    \displaystyle 
    \qty[ z z_0 e^{- 2 \omega \qty( N - N_0 ) } ]^{-(2 \ell + 1)/4} 
    }{
    1 - e^{- 2 \omega \qty( N - N_0 ) }
    } 
    \notag \\ 
    &\qquad\quad \times \exp \qty[- \qty( z + z_0 ) \frac{ e^{- 2 \omega \qty( N - N_0 ) } }{1 - e^{- 2 \omega \qty( N - N_0 ) }} ] 
    I_{\ell + 1/2}
    \qty(
    \frac{\sqrt{ z z_0 }}{\sinh \qty[ \omega \qty( N - N_0 ) ] } 
    ) 
    \,\, , 
    \label{eq:exs_rho_dist}
\end{align}
where $z \coloneqq y^{2}$. 
As $N \to \infty$, this expression asymptotes to the stationary distribution together with the stationary statistical moments, 
\begin{subequations}
\label{eq:exs_stdist_rho}
\begin{align}
	f_{\infty} (\phi) 
    &= \qty( \alpha \sqrt{ \frac{ 8 \pi^{2} }{ H^{2} } } \, ) \frac{2}{\Gamma \qty( \ell + 3/2 )} z^{\ell +1} e^{-z}
	\,\, , 
    \\ 
    \expval{ \phi^{n} }_{\infty} 
    &= \qty( \alpha \sqrt{ \frac{ 8 \pi^{2} }{ H^{2} } } \, )^{- n} 
    \frac{ \Gamma [\ell + (n+3) / 2 ] }{ \Gamma (\ell + 3/2) } 
    \,\, . 
\end{align}
\end{subequations}

\paragraph{\textit{Trigonometric Scarf}}
Aside from (semi)infinite-domain potentials, there are several periodic potentials that can be solved exactly in terms of the Jacobi polynomial. 
Of all of them, the trigonometric Scarf potential (also called the Scarf I potential) gives the simplest case. 
The potential of the test field is given by $v ( y ) = - s \ln \cos y + t \ln \qty{ \qty[ 1 - \tan (y/2) ] / \qty[ 1 + \tan (y/2) ] }$ where $- \pi / 2 < y < \pi / 2$. 
The eigenenergy depends on the level index $n$ non-linearly, which prohibits us from obtaining a closed-form expression in stochastic inflation. 
Though the full expression can be found in Ref.~\cite{Honda:2024evc}, let us restrict ourselves to the stationary state. 
In that case, the distribution and correlation functions greatly simplify to be 
\begin{subequations}
    \label{eq:exs_stdist_sc1_m}
    \begin{align}
        f_{\infty} (\phi) 
        &= \qty( \alpha \sqrt{ \frac{ 8 \pi^{2} }{ H^{2} } } \, ) 
        \times 
        \frac{2s}{2^{2s}} \frac{ \Gamma (2s) }{ \Gamma (s - t + 1/2) \Gamma (s + t + 1/2) } (1 - z)^{s - t} (1 + z)^{s + t}
        \,\, , 
        \label{eq:exs_stdist_sc1}
        \\ 
        \expval{ \phi^{n} }_{\infty} 
        &= \qty( \alpha \sqrt{ \frac{ 8 \pi^{2} }{ H^{2} } } \, )^{- n} 
        \frac{2 s}{2^{2 s}} 
        \frac{ \Gamma (2 s) }{ \Gamma (s - t + 1/2) \Gamma (s + t + 1/2) } 
        \int_{- \pi / 2}^{\pi / 2} \dd y \, y^{n} \cos^{2 s} y \qty( \frac{1 + \sin y}{1 - \sin y} )^{t} 
        \,\, . 
        \label{eq:exs_stdist_sc1_c}
    \end{align}
\end{subequations}
where $z \coloneqq \sin y$ and $\Gamma (z)$ is the Gamma function (generalised factorial). 
As will be seen in the next section, one of the parameters $s$ together with $\alpha$ can be identified with the mass of the test field, while $t$ measures the asymmetry of the potential around the global minimum located at $\widetilde{y} = \mathrm{arcsin} (t/s)$. 

\paragraph{\textit{Trigonometric Rosen--Morse}}
The last exactly solvable situation that consists only of the discrete eigenstates is a test field in the potential $v (y) = - (t/s) y - s \ln \cos y $, confined in the region $- \pi / 2 < y < \pi / 2$. 
It corresponds to the trigonometric Rosen--Morse (Rosen--Morse I) potential in quantum mechanics. 
The full expression taking into account the time evolution can again be found in Ref.~\cite{Honda:2024evc}. 
In the late-time limit, it reduces to the stationary distribution in terms of $z \coloneqq \tan y$, given by 
\begin{equation}
	f_{\infty} (\phi) 
        = \qty( \alpha \sqrt{ \frac{8 \pi^{2}}{H^{2}} } \, ) \frac{2^{2s}}{2 \pi} \frac{ \Gamma (s + 1 - i t/ s) \Gamma (s + 1 + i t / s) }{ s \Gamma ( 2 s ) } 
        \frac{1}{(1 + z^{2})^{s}} 
        \exp \qty( \frac{2 t}{s} \arctan z )  
    \label{eq:exs_stdist_rm1}
\end{equation}
The reality $f_{\infty} (\phi) \in \mathbb{R}$ can immediately be confirmed by virtue of the identity $\overline{ \Gamma (\bullet) } = \Gamma ( \, \overline{\bullet} \, )$, where an overline denotes the complex conjugate. 

\vspace{1.0\baselineskip}

As will be seen in the next section, all four potentials of the test field start with the quadratic term around each global minimum. 
This enables us to identify the model parameters $\alpha$ and $s$ with the mass of the test field. 
One can then make probabilistic arguments by studying the structure of the parameter space, taking into account the observed amplitude of the curvature perturbation and assuming that it originated from a curvaton. 

\section{Application to a curvaton scenario}
\label{sec:curvaton}

Having prepared several exact distribution functions and statistical moments of a test field in stochastic inflation, in this section those will be applied to a curvaton scenario to derive the analytical distribution of the curvature perturbation in each case, to investigate the structure of the parameter space. 

A curvaton~\cite{Lyth:2001nq, Moroi:2001ct, Enqvist:2001zp, Lyth:2002my} can be regarded as a test field. 
It can constitute a part of or all the observed curvature perturbation. 
During or after inflation depending on the mass and decay rate, the fluctuation of the curvaton is eventually transformed into the curvature perturbation. 
This enables us to relate the curvaton field $\phi$ to the curvature perturbation $\zeta$.
The model parameters therefore enjoy several constraints from observation, such as the amplitude of the total curvature perturbation or its local-type non-Gaussianity parametrised by $f_{\rm NL}$. 
Accordingly, the parameter space can be divided into several regions based on a criterion that distinguishes whether a set of the parameters may or may not give a non-negligible contribution to the curvature perturbation. 
In particular, since the stochastic behaviour of the curvaton and thus curvature perturbation are described by the statistical quantities presented in Ref.~\cite{Honda:2024evc} and reviewed in section~\ref{subsec:exactsol}, probabilistic statements for the model parameters can be made. 
For instance, one can investigate in which region in the whole parameter space and to what extent a non-negligible fraction of the observed total curvature perturbation can originate from the curvaton. 
Those questions are exactly what will be addressed in section~\ref{subsec:curvaton_param}. 
To do this, let us start by analytically deriving the distribution of the curvature perturbation converted from $f_{\infty} (\phi)$ under the relation between $\phi$ and $\zeta$, by identifying the test field considered in section~\ref{sec:stochinf} with the curvaton. 

\subsection{From curvaton to curvature perturbations}
\label{subsec:curv_func}

The conversion of the curvaton into the curvature perturbation enables us to have a functional relation between them. 
When the potential of the curvaton is purely quadratic, the relation is given by~\cite{Lyth:2009imm} 
\begin{equation}
	\zeta = \frac{ r_{\rm decay} }{3 \pi} \frac{H}{\phi} 
	\,\, . 
	\label{eq:curvaton_relori}
\end{equation}
In Eq.~(\ref{eq:curvaton_relori}), the curvaton field and the Hubble parameter are implicitly evaluated at the horizon exit of a relevant mode. 
It should be noted that $\zeta$ in Eq.~(\ref{eq:curvaton_relori}) is a part of the total curvature perturbation unless all of the total originates from the curvaton. 
In general, the observed, and hence total, curvature perturbation consists of the two contributions from both inflaton and curvaton, as $\zeta_{\rm obs} = \zeta_{\rm infl} + \zeta_{\rm curv}$, where $\zeta_{\rm infl}$ originates from the inflaton while $\zeta_{\rm curv}$ does from the curvaton given by Eq.~(\ref{eq:curvaton_relori}). 
While Eq.~(\ref{eq:curvaton_relori}) ceases to be an exact relation (at the leading order of fluctuations) when the curvaton is in a non-quadratic potential such as the ones considered in section~\ref{subsec:exactsol}, regarding it as our first-step approximation to demonstrate usage of the exact solutions in stochastic inflation, the relation (\ref{eq:curvaton_relori}) will be employed in the following. 
It should be noted, however, that the deviation from the exact relation (\ref{eq:curvaton_relori}) for a non-quadratic potential can be relevant~\cite{Enqvist:2005pg, Enqvist:2008gk, Huang:2008zj, Huang:2008bg, Enqvist:2009zf, Enqvist:2009ww, Enqvist_2010, Byrnes:2011gh, Kawasaki:2011pd, 10.1143/PTPS.190.62}.  To avoid making the following discussion complicated, studying such effects is left for future work. 

Whether the curvaton does or does not dominate the universe at its decay is relevant to the generated curvature perturbation. 
This is why the weighted energy-density fraction of the curvaton defined by 
\begin{equation}
	r_{\rm decay} 
	\coloneqq \eval{ \frac{ 3 \rho_{\phi} }{ 3 \rho_{\phi} + 4 \rho_{\rm r} } }_{t = t_{\rm decay}} \,\, , 
\end{equation}
appears in Eq.~(\ref{eq:curvaton_relori}). 
Here, $\rho_{\phi}$ and $\rho_{\rm r}$ are the energy density of the curvaton and radiation, respectively, both of which are in general functions of time. 
The initial conditions at the onset of the reheating are given by $\rho_{\rm rad} \simeq 3 M_{\rm P}^{2} H^{2}$ and $\rho_{\phi} \simeq V (\phi)$. 
Given that matter and radiation decay according to $1 / a^{3} (t)$ and $1 / a^{4} (t)$ respectively, assuming the radiation-domination, and evaluating those quantities at the decay time, it reads~\cite{Enqvist:2012xn, Lerner_2014} 
\begin{equation}
    r_{\rm decay} 
    \simeq \frac{ (\phi / M_{\rm P})^{2} }{ (\phi / M_{\rm P})^{2} + 4 (\Gamma / m)^{1/2} } 
    \qquad 
    \text{and}
    \qquad 
    \zeta 
	\simeq \frac{1}{3 \pi} \frac{ (H / M_{\rm P}) (\phi / M_{\rm P}) }{ (\phi / M_{\rm P})^{2} + 4 (\Gamma / m)^{1/2} } 
	\,\, .  
	\label{eq:ph_curv_rel}
\end{equation}
The $\Gamma$ in those equations is the decay rate of the curvaton. 
In general, it is model-dependent and can depend on the mass of the curvaton and/or the field value (the latter case is called modulated reheating~\cite{Kofman:2003nx, PhysRevD.69.023505, PhysRevD.73.103516, PhysRevD.69.043508, PhysRevD.72.023512, PhysRevD.77.023505, PhysRevD.78.063545, Choi:2012te, Langlois:2013dh, Assadullahi:2013ey, Enomoto:2013qf, RevModPhys.78.537}). 
Here, however, $\Gamma$ is assumed to be rather a constant and a free parameter. 
The sudden-decay approximation is also assumed, that is, the curvaton instantaneously decays when $H = \Gamma$~\cite{Lyth:2001nq, Lyth:2005fi, Sasaki:2006kq, Malik:2006pm, Kitajima:2014xna}. 
The energy density of the curvaton field is negligible during inflation, but the curvaton starts to oscillate when its mass and the Hubble scale coincide, $H = m$, after which the curvaton may dominate the universe since $\rho_{\phi} / \rho_{\rm rad} \propto a$. 
This domination can in particular be realised if $r_{\rm decay} \simeq 1$, while the curvaton does not dominate the universe if $r_{\rm decay} \ll 1$. 
In the latter case, a large non-Gaussianity may instead be realised because of the relation $f_{\rm NL} \approx 1 / r_{\rm decay}$. 
In both cases, however, there exist regions in the parameter space with which the curvaton can give rise to a non-negligible fraction of the total curvature perturbation, as will be probabilistically confirmed in section~\ref{subsec:curvaton_param}. 

\begin{figure}
	\centering
	\begin{subfigure}[b]{0.995\textwidth}
        \centering
   		\includegraphics[width=0.5\linewidth]{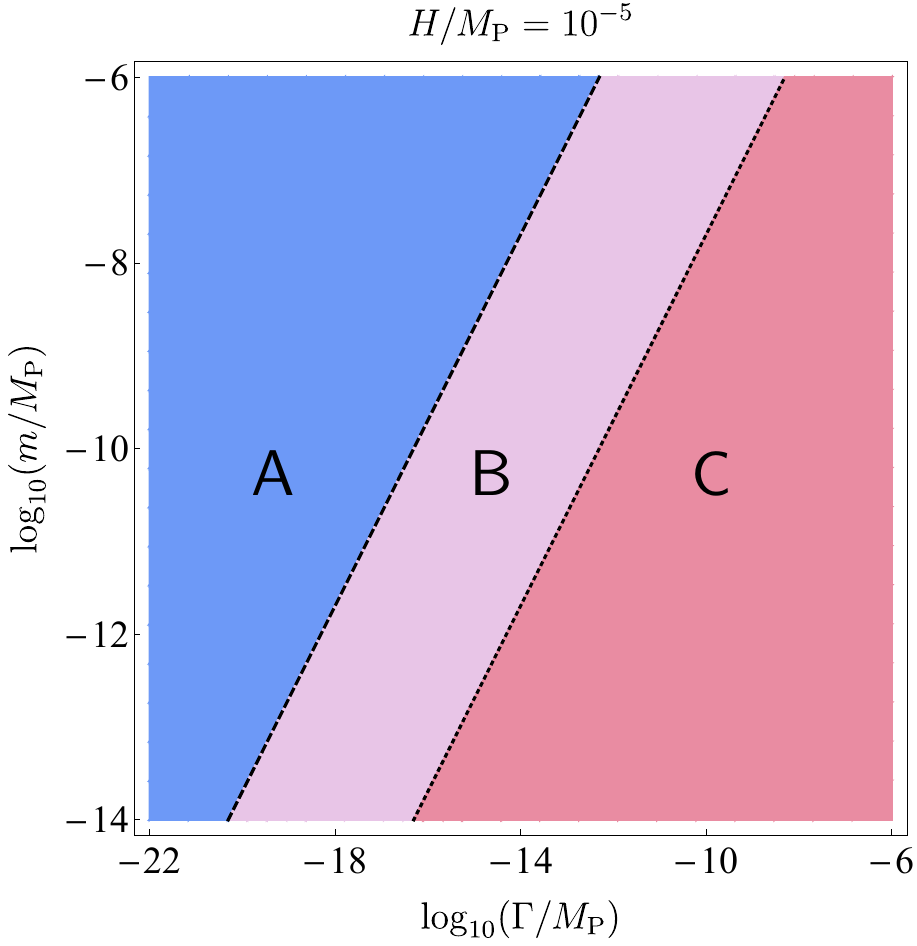}
   		\caption{
			The parameter space typically divided into several regions. 
		}
   		\label{fig:prst_a}
	\end{subfigure}
    \\[4.0ex]
    \begin{subfigure}[b]{0.995\textwidth}
        \centering
        \includegraphics[width=0.65\linewidth]{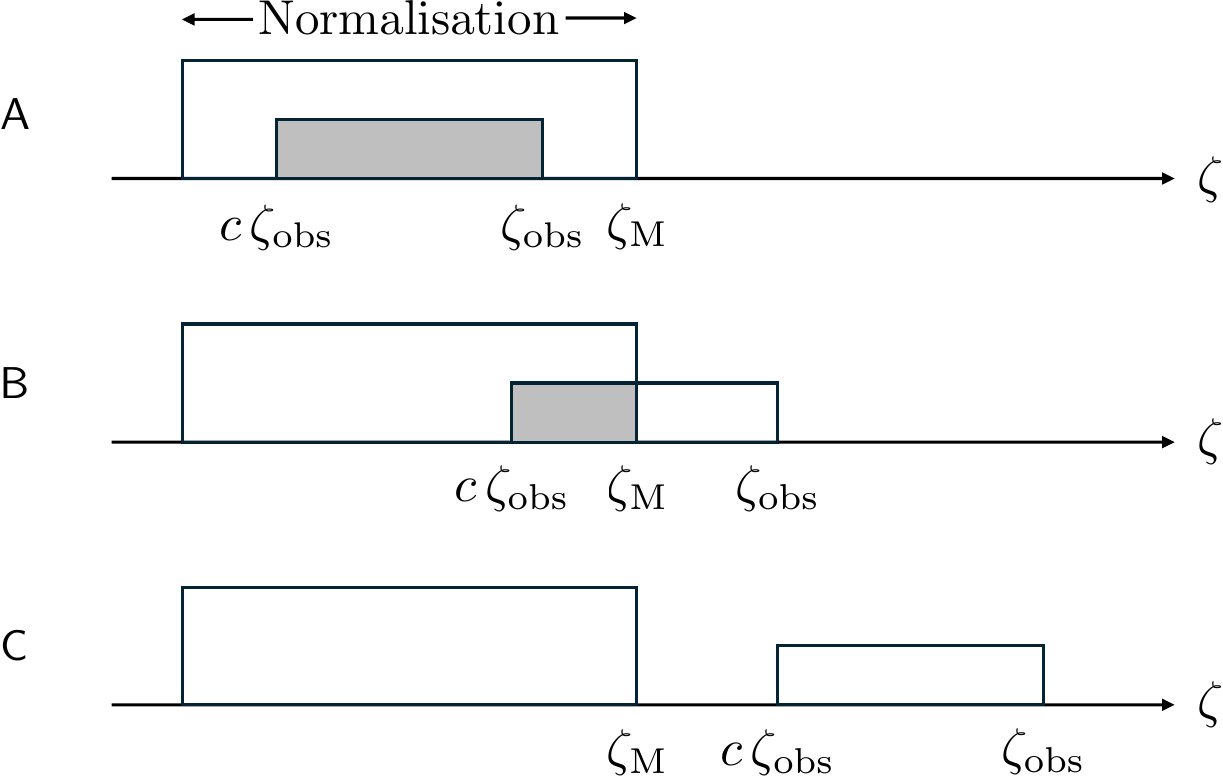}
   		\caption{
			The interplay between the domain to be normalised and integrated for the probability. 
		}
   		\label{fig:prst_b}
    \end{subfigure}
    \caption{
        (a) 
        The parameter space of the curvaton model considered in the main text is divided into several regions. 
        A parameter set $(\Gamma, \, m)$ in the region $\mathsf{A}$ guarantees at least $10\%$ of, and can give rise to, the observed total curvature perturbation, hence will be focussed on in this paper. 
        Each pair of the two neighbouring regions is divided by the straight line given either by Eq.~(\ref{eq:pmt_1}) or (\ref{eq:pmt_2}). 
        (b) 
        The relation amongst $c \, \zeta_{\rm obs}$, $\zeta_{\rm obs}$, and $\zeta_{\rm M}$ in each region. 
        The grey shaded domain for the region $\mathsf{A}$ is integrated to define the probability $\mathbb{P} (c \, \zeta_{\rm obs} < \zeta < \zeta_{\rm obs})$ in section~\ref{subsec:curvaton_param}. 
        The lower bound of the domain to be normalised can either be $0$ (for the harmonic potentials) or $[2 x / (1 + x^{2})] \zeta_{\rm M}$ (for the trigonometric potentials) with $x$ being defined in Eq.~(\ref{eq:curv_tsc_avreg}).
        }
    \label{fig:prst}
\end{figure}

The relation (\ref{eq:ph_curv_rel}) is a quadratic equation in $\phi$ and can be solved inversely, 
\begin{equation}
	\phi_{\pm}
	= \frac{H}{6 \pi \zeta} \qty[ 
		1 \pm \sqrt{ 1 - \qty( \frac{\zeta}{\zeta_{\rm M}} )^{2} } 
		\, 
	] \,\, , 
	\qquad 
	\zeta_{\rm M}^{2} 
	\coloneqq 
	\frac{ (H / M_{\rm P})^2 }{ (12 \pi)^{2} ( \Gamma / m )^{1/2} } 
	\,\, . 
	\label{eq:ph_curv_rel_inv}
\end{equation}
The ``$+$'' branch corresponds to the situation where $r_{\rm decay} \simeq 1$, while the ``$-$'' branch does to $r_{\rm decay} < 1$. 
From Eq.~(\ref{eq:ph_curv_rel_inv}), it can be seen that the curvature perturbation originating from the curvaton cannot be enhanced arbitrarily. 
Rather, there exists an upper bound determined by the maximum value $\zeta_{\rm M}$. 
The model-parameter dependence propagates also through $\zeta_{\rm M}$. 
When the observed value of the curvature perturbation will be taken into account in section~\ref{subsec:curvaton_param}, the order between $\zeta_{\rm obs}$ and $\zeta_{\rm M}$ becomes relevant. 
This is why, as Figure~\ref{fig:prst_a} shows, the parameter space is typically divided into several regions. 
For the fraction of the total curvature perturbation, $0 < c < 1$, these blue ($\mathsf{A}$), thin-magenta ($\mathsf{B}$), and red ($\mathsf{C}$) regions are defined by $\zeta_{\rm obs} < \zeta_{\rm M}$, $c \, \zeta_{\rm obs} < \zeta_{\rm M} < \zeta_{\rm obs}$, and $\zeta_{\rm M} < c \, \zeta_{\rm obs}$, respectively. 
The dashed and dotted lines in Figure~\ref{fig:prst_a} are therefore given respectively by (hereafter $c = 0.1$ and $\zeta_{\rm obs} = 10^{-5}$ are fixed for concreteness) 
\begin{subequations}
    \begin{align}
        \log_{10} \qty( \frac{m}{M_{\rm P}} ) 
        &= \log_{10} \qty( \frac{\Gamma}{M_{\rm P}} ) + 4 \log_{10} (\zeta_{\rm obs}) - 4 \log_{10} \qty( \frac{H}{M_{\rm P}} ) + 4\log_{10} (12 \pi) 
        \,\, , 
        \label{eq:pmt_1}
        \\ 
        \log_{10} \qty( \frac{m}{M_{\rm P}} ) 
        &= \log_{10} \qty( \frac{\Gamma}{M_{\rm P}} ) + 4 \log_{10} (\zeta_{\rm obs}) - 4 \log_{10} \qty( \frac{H}{M_{\rm P}} ) + 4\log_{10} (12 \pi) + 4 \log_{10} c 
        \,\, .  
        \label{eq:pmt_2}
    \end{align}
\end{subequations}
In particular, for a set of the model parameters in a subset of $\mathsf{A}$, the full amount of the observed curvature perturbation may be explained solely by the curvaton (with a specific probability). 
Also, it is said that one can realise a ``non-negligible'' contribution from the curvaton when $c \, \zeta_{\rm obs} < \zeta_{\rm M}$ is satisfied, meaning that at least $10\%$ of the total can originate from the curvaton (with a specific probability). 
The set of the parameters $(\Gamma, \, m)$ in the region $\mathsf{C}$, on the other hand, cannot achieve a non-negligible contribution. 
Only less than $10\%$ of the total curvature perturbation can originate from the curvaton no matter how one searches in $\mathsf{C}$. 
The parameters in the region $\mathsf{B}$ can result in a non-negligible contribution, but another source would be needed to explain the total. 
The regions $\mathsf{B}$ and $\mathsf{C}$ will therefore not be of our interest hereafter. 
When the parameter space will be analysed in section~\ref{subsec:curvaton_param}, the region where the inequality $c \, \zeta_{\rm obs} < \zeta < \zeta_{\rm obs}$ holds rather will be focussed on.\footnote{
    A short discussion about the parameter-space region $\mathsf{B}$ in addition to region $\mathsf{A}$ can be found in Appendix~\ref{appx:appx1}. 
}
It is nothing but the region $\mathsf{A}$ when the parameter-space regions are divided according to Figure~\ref{fig:prst_a}. 

For the curvaton in the harmonic potentials including the radial-harmonic-oscillator case in section~\ref{subsec:exactsol}, the above classification of the parameter space (Figure~\ref{fig:prst_a}) applies and the domain of $\zeta$ to define the probability is summarised in Figure~\ref{fig:prst_b}. 
On the other hand, the parameter space is endowed with a richer structure than Figure~\ref{fig:prst_a} for the two trigonometric cases and careful arguments are needed. 
There, Figure~\ref{fig:prst_tri} and Table~\ref{tab:cbc} will be the counterpart of Figure~\ref{fig:prst_a} and Figure~\ref{fig:prst_b}, respectively. 
While the normalisation of the distribution (as shown in Figure~\ref{fig:prst_b}) is necessary to calculate the probability for all the cases considered below, the lower bound of the domain to be normalised is not shown in Figure~\ref{fig:prst_b} since it depends on the potential of the curvaton, see section~\ref{subsec:curvaton_param} for the details. 

\subsection{Distribution of curvature perturbations}
\label{subsec:curv_dist}

The distribution function of the curvature perturbation, denoted by $\mathbb{P} (\zeta)$ hereafter, can be obtained through conservation of probability under the relation (\ref{eq:ph_curv_rel}), or equivalently (\ref{eq:ph_curv_rel_inv}). 
Assuming the stationary state as announced below Eq.~(\ref{eq:exs_stdist_ho}), for the stationary distribution of the curvaton identified with either of Eqs.~(\ref{eq:exs_stdist_ho}), (\ref{eq:exs_stdist_rho}), (\ref{eq:exs_stdist_sc1}), or (\ref{eq:exs_stdist_rm1}), $\mathbb{P} (\zeta)$ is given by 
\begin{equation}
	\mathbb{P} ( \zeta )
	= \sum_{\lambda = \pm} f_{\infty} (\phi_{\lambda}) \, \abs{ \dv{\phi}{\zeta} } 
	= f_{\infty} (\phi_{+}) \, \abs{ \dv{\phi_{+}}{\zeta} } 
	+ f_{\infty} (\phi_{-}) \, \abs{ \dv{\phi_{-}}{\zeta} } 
	\,\, .  
	\label{eq:curvaton_transf}
\end{equation}
The Jacobian associated with the variable transformation from $\phi$ to $\zeta$ is given by, introducing the auxiliary variable $\mathsf{Y}$ (that satisfies $0 < \mathsf{Y} < 1$), 
\begin{equation}
	\dv{\phi_{\pm}}{\zeta} 
	= \frac{H}{6 \pi \zeta_{\rm M}^{2}} \frac{1}{\mathsf{Y} (1 \mp \mathsf{Y})} 
    > 0 
	\,\, , 
	\qquad 
	\mathsf{Y} = \mathsf{Y} (\zeta) 
	\coloneqq \sqrt{1 - \qty( \frac{\zeta}{\zeta_{\rm M}} )^{2}} 
	\,\, . 
	\label{eq:curvaton_fny}
\end{equation}
In what follows, this new variable $\mathsf{Y}$ will exclusively be used instead of $\zeta$ itself. 
Note again that the dependence on the model parameters propagates through $\mathsf{Y}$ since it contains $\zeta_{\rm M}$. 
The relation~(\ref{eq:ph_curv_rel_inv}) expressed in terms of $\mathsf{Y}$ is  
\begin{equation}
	\frac{\phi_{\pm}}{H} 
	= \frac{1}{6 \pi \zeta_{\rm M}} \frac{1 \pm \mathsf{Y}}{ \sqrt{1 - \mathsf{Y}^{2}} } 
	= \frac{1}{6 \pi \zeta_{\rm M}} \sqrt{ \frac{1 \pm \mathsf{Y}}{1 \mp \mathsf{Y}} } 
	\,\, , 
 \label{eq:curv_rel_inty}
\end{equation}  
and the distribution of the curvature perturbation (\ref{eq:curvaton_transf}) can also be written in terms of $\mathsf{Y}$, 
\begin{equation}
	\mathbb{P} (\zeta) 
	= \qty( \frac{\sqrt{2} \, \alpha}{3 \zeta_{\rm M}^{2}} ) 
	\frac{1}{\mathsf{Y}} 
	\qty[ 
		\frac{1}{1-\mathsf{Y}} \frac{ f_{\infty} (\phi_{+}) }{ \alpha \sqrt{ 8 \pi^{2} / H^{2} } } 
		+ \frac{1}{1+\mathsf{Y}} \frac{ f_{\infty} (\phi_{-}) }{ \alpha \sqrt{ 8 \pi^{2} / H^{2} } } 
	] 
	\,\, , 
	\label{eq:curvaton_dist_master}
\end{equation}
where it is understood that $\phi_{\pm}$ in the arguments are replaced in terms of $\mathsf{Y}$ using Eq.~(\ref{eq:curv_rel_inty}). 
The distribution (\ref{eq:curvaton_dist_master}) gives that of the curvature perturbation converted from the curvaton, assuming that the duration of inflation is sufficiently long so that the universe is already equilibrated when the distribution of the curvaton is calculated. 
Since the curvature perturbation inherits the dynamics of the curvaton, its distribution depends on the curvaton model. 
In other words, the model parameters, $\Gamma$ and $m$ in $\zeta_{\rm M}$ and $\mathsf{Y}$, as well as the potential on which the curvaton rolls down, can probabilistically be constrained by taking the observed amount of the curvature perturbation into account. 

In Eq.~(\ref{eq:curvaton_dist_master}), $\alpha \sqrt{ 8 \pi^{2} / H^{2} }$ is factored out from $f_{\infty} (\phi)$ since the distribution function of $\phi$ is proportional to the factor in every case, see section~\ref{subsec:exactsol}. 
The distribution (\ref{eq:curvaton_dist_master}) must be properly normalised according to 
\begin{equation}
	\int \dd \zeta \, \mathbb{P} (\zeta) = 1 
	\,\, . 
\end{equation}
The domain to be integrated depends on the potential of the curvaton, as will be discussed in section~\ref{subsec:curvaton_param}. 

\subsection{Promising regions in parameter space}
\label{subsec:curvaton_param}

Once the domain to be normalised is fixed, the cumulative distribution can be calculated by integrating $\mathbb{P} (\zeta)$ in a range, say $c \, \zeta_{\rm obs} < \zeta < \zeta_{\rm obs}$ for the parameters ($\Gamma, \, m)$ in the region $\mathsf{A}$ in Figure~\ref{fig:prst_a}. 
It is this cumulative distribution that measures the probability to realise the non-negligible contribution of the curvaton to the observed curvature perturbation, relating it to the model parameters, the decay rate and the mass. 
To put it more concretely, let us define the cumulative distribution in the standard manner, 
\begin{equation}
	\mathbb{P} (\zeta_{1} < \zeta < \zeta_{2}) 
	\coloneqq 
	\int_{\zeta_{1}}^{\zeta_{2}} \, \dd \zeta \, \mathbb{P} (\zeta) 
	\,\, . 
	\label{eq:curvaton_cumul}
\end{equation}
The integrand is assumed to be properly normalised in Eq.~(\ref{eq:curvaton_cumul}). 
Now, having the region $\mathsf{A}$ in mind as an example, let us explain the usage of Eq.~(\ref{eq:curvaton_cumul}). 
With $\zeta_{\rm obs} < \zeta_{\rm M}$ being satisfied, and if the two limits of the integral are fixed to be $\zeta_{1} = c \, \zeta_{\rm obs}$ and $\zeta_{2} = \zeta_{\rm obs}$, the cumulative distribution~(\ref{eq:curvaton_cumul}) gives the probability that at least $100 c \% = 10\%$ of the total $\zeta_{\rm obs}$ originates from the curvaton, while the contribution of the curvaton does not exceed $100\%$ of the total $\zeta_{\rm obs}$. 
As it will be seen below, a subset of $\mathsf{A}$ can give rise to such a non-negligible amount of curvature perturbation with a high probability. 
For other regions where $\mathbb{P} (\zeta_{1} < \zeta < \zeta_{2})$ is small, on the other hand, the complementary quantity $1 - \mathbb{P} (\zeta_{1} < \zeta < \zeta_{2})$ is rather large, which means that with a high probability the contribution of the curvaton may fall short of $10\%$ of or excess the total $\zeta_{\rm obs}$. 
Such a combination of the decay rate $\Gamma$ and the mass $m$ are therefore probabilistically unlikely to explain the observations. 
To summarise, as long as region $\mathsf{A}$ is considered, our statement would be, for instance, that a set of the parameters $(\Gamma, \, m)$ can generate a non-negligible fraction (including the total) of the total curvature perturbation. 
Every statement must be associated with a specific probability given by $p\% = 100 \mathbb{P} (\zeta_{1} < \zeta < \zeta_{2})\%$. 

In what follows, each case where the curvaton in a harmonic-type potential and in a trigonometric-type potential will separately be studied. 
This is because the radial harmonic oscillator includes the pure harmonic oscillator in the limit $\ell \to -1$ (in some sense), and the trigonometric Rosen--Morse potential and the trigonometric Scarf potential become almost the same in the symmetric limit $t \to 0$, and because the parameter-space structure differs from each other due to the presence of the boundary at the finite location in the latter case (recall that $y = \alpha \sqrt{ 8 \pi^{2} / H^{2} } \, \phi$ has its boundaries at $y = \pm \pi / 2$). 

\subsubsection{Curvaton in harmonic potential}
\label{subsec:curvaton_harmonic}

Let us start with the curvaton in the pure harmonic potential, $V (\phi) / H^{4} = (3 / 4 \pi^{2}) [ v (y) - v (\widetilde{y}) ]$ where $v (y) = y^{2} / 2$, $\widetilde{y} = 0$, and $y = \alpha \sqrt{8 \pi^{2} / H^{2} } \, \phi$. 
The parameter $\alpha$ is identified with the mass of the curvaton according to $\alpha = (m / H) / \sqrt{6}$, for which the mass-term potential $V (\phi) = m^{2} \phi^{2} / 2$ is recovered. 
This identification is consistent with Ref.~\cite{Hardwick:2017fjo} (see also the mean and variance of $\phi$ given in Eqs.~(\ref{eq:exs_mv_a}) and (\ref{eq:exs_mv_b}), the coefficients in the exponents of which, $1 / 2 \alpha^{2}$ and $1 / 4 \alpha^{2}$, are called the relaxation and decoherence timescales respectively in Ref.~\cite{Enqvist:2012xn}). 
With this identification, and by assuming that the curvaton has already settled down in the stationary state as announced previously, the distribution of the curvature perturbation follows from Eq.~(\ref{eq:exs_stdist_ho}) and our master formula (\ref{eq:curvaton_dist_master}) that
\begin{equation}
	\mathbb{P} (\zeta) 
	= 
	\frac{2}{\sqrt{\pi}} \, 
	\qty( \frac{\sqrt{2} \, \alpha}{3 \zeta_{\rm M}^{2}} ) \frac{1}{\mathsf{Y}} 
	\qty{ 
		\frac{1}{1 - \mathsf{Y}} \exp \qty[ 
			- \qty( \frac{\sqrt{2} \, \alpha}{3 \zeta_{\rm M}} )^{2} \frac{ 1 + \mathsf{Y} }{ 1 - \mathsf{Y} } 
		] + \frac{1}{1 + \mathsf{Y}} \exp \qty[ 
			- \qty( \frac{\sqrt{2} \, \alpha}{3  \zeta_{\rm M}} )^{2} \frac{ 1 - \mathsf{Y} }{ 1 + \mathsf{Y} } 
		] 
	} 
	\,\, . 
	\label{eq:curv_dist_ho}
\end{equation}
Note that the model-parameter dependence propagates through not only $\mathsf{Y}$, but now also through $\alpha$. 
Note also that the dependence on $\zeta$ in $\mathbb{P} (\zeta)$ appears only in the quadratic form through $\mathsf{Y} = \sqrt{1 - ( \zeta / \zeta_{\rm M})^{2}}$, see Eqs.~(\ref{eq:curvaton_fny}) and (\ref{eq:curvaton_dist_master}). 
The  following normalisation condition for the distribution of the curvature perturbation is imposed, 
\begin{equation}
	\int_{0}^{\zeta_{\rm M}} \dd \zeta \, \mathbb{P} (\zeta) = 1 
	\,\, . 
 \label{eq:curv_nom_ho}
\end{equation}
The normalisation (\ref{eq:curv_nom_ho}) explains the additional factor two in Eq.~(\ref{eq:curv_dist_ho}), which is absent in Eq.~(\ref{eq:exs_stdist_ho}). 

\begin{figure}
    \centering
    \includegraphics[width=.495\textwidth]{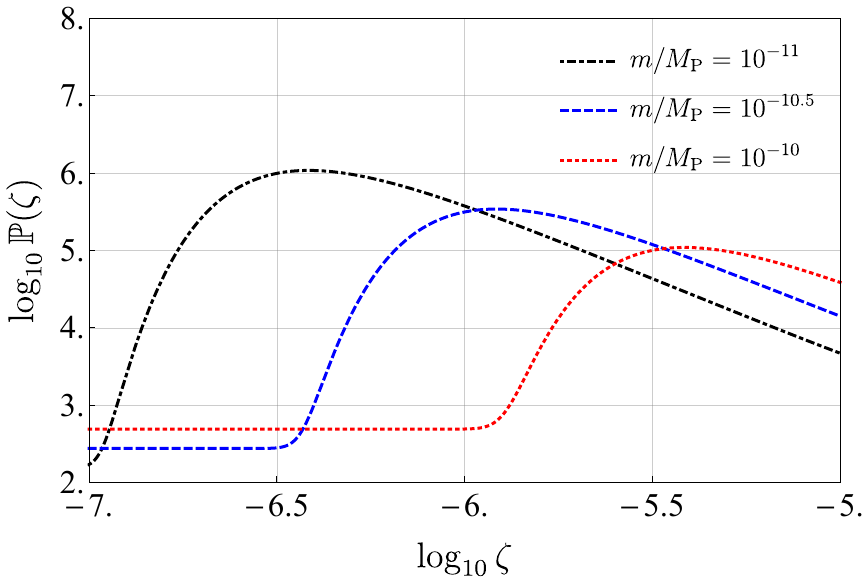}
    \hfill
    \includegraphics[width=.495\textwidth]{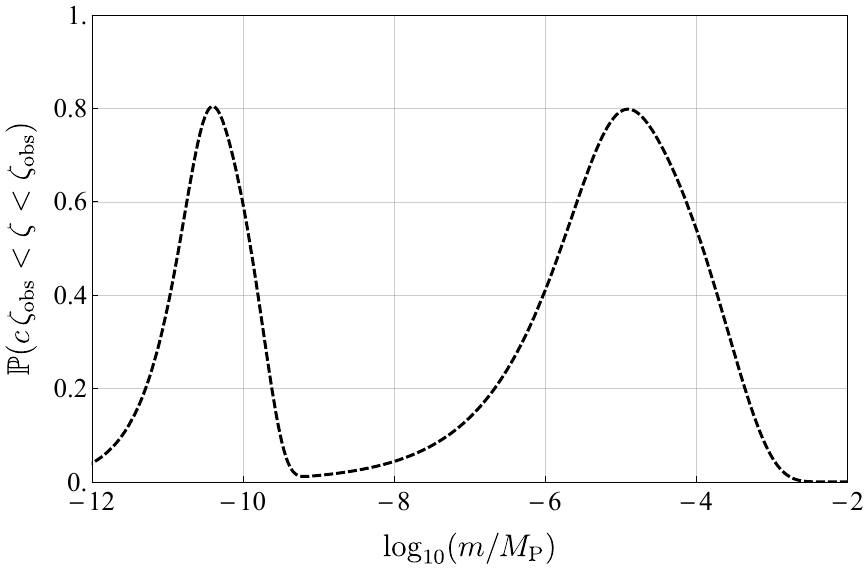}
    \caption{
        (\textit{Left}) 
        The distribution of the curvature perturbation originating from the curvaton in the potential $v (y) = y^{2} / 2$. 
        (\textit{Right}) 
        The cumulative distribution of the curvature perturbation for each mass of the curvaton. 
    }
    \label{fig:curvaton_dist_ho}
\end{figure}

The left panel of Figure~\ref{fig:curvaton_dist_ho} shows the distribution of the curvature perturbation given by Eq.~(\ref{eq:curv_dist_ho}), in the range from $\zeta = 10^{-7}$ to $10^{-5}$. 
For the concrete set of the parameters, $\Gamma / M_{\rm P} = 10^{-19}$ and $H / M_{\rm P} = 10^{-5}$ that belongs to the region $\mathsf{A}$ in Figure~\ref{fig:prst_a} are implemented. 
It can be seen that both the location and amplitude at the peak depend on the mass of the curvaton, which implies that only a specific choice of mass can give rise to the high probability of realising the contribution of the curvaton. 
When the distribution is integrated from $\zeta_{1} = 0.1 \, \zeta_{\rm obs}$ to $\zeta_{2} = \zeta_{\rm obs} < \zeta_{\rm M}$, it defines the probability that at least $10\%$ of the total curvature perturbation can be explained by the curvaton. 
The right panel in Figure~\ref{fig:curvaton_dist_ho} shows the cumulative distribution defined in Eq.~(\ref{eq:curvaton_cumul}) for Eq.~(\ref{eq:curv_dist_ho}), integrated in the domain $0.1 \, \zeta_{\rm obs} < \zeta < \zeta_{\rm obs}$. 
The double peaks can be found, one of which corresponds to the smaller mass that comes from the positive branch of the solution, implying that the curvaton dominates the universe at its decay.  
On the other hand, the larger-mass peak corresponds to the negative branch with $r_{\rm decay} < 1$, though for the chosen parameters the mass of the curvaton and the inflationary energy scale are compatible around the peak. 
For intermediate masses, the peak of $\mathbb{P} (\zeta)$ does not come inside the domain to be integrated, resulting in smaller probabilities. 
The plot is qualitatively consistent with the result given in Figure 3 in Ref.~\cite{Lerner_2014}, although here the different values of the parameters are implemented for an illustrative purpose, and the effect that comes from $f_{\rm NL}$ was also taken into account in addition to $\zeta_{1} < \zeta < \zeta_{2}$ in the same reference, but not here. 

Seeing the right panel of Figure~\ref{fig:curvaton_dist_ho} in more detail, the two locations where the cumulative distribution becomes extremum at $m / M_{\rm P} \sim 10^{-10}$ and $m / M_{\rm P} \sim 10^{-5}$. 
These ``preferred'' masses can also be found by an easier order estimation. 
Since, from Eq.~(\ref{eq:exs_mv}), in the stationary (equilibrium) state the field typically deviates from the bottom by $\expval{ ( \phi / H )^{2} }_{\infty}^{1/2} = \sqrt{ 3 / 8 \pi^{2} } \,  ( H / m ) \sim ( H / m )$, the generated curvature perturbation (\ref{eq:curvaton_relori}) is of order of $\zeta \sim r_{\rm decay} ( m / H )$. 
When the curvaton dominates the universe at its decay with $r_{\rm decay} \simeq 1$, the amplitude is simply determined by the mass of the curvaton for a given energy scale, regardless of the decay rate. 
In the case of Figure~\ref{fig:curvaton_dist_ho}, one obtains $\zeta \sim 10^{-5}$ for $m / M_{\rm P} \sim 10^{-10}$ when $H / M_{\rm P} \sim 10^{-5}$. 
For $r_{\rm decay} \simeq (\phi / M_{\rm P})^{2} \sqrt{ m / \Gamma }$, on the other hand, the amplitude of the produced curvature perturbation is estimated as $\zeta \sim (H / M_{\rm P})^{3} (\Gamma / M_{\rm P} )^{-1/2} (m / M_{\rm P})^{-1/2}$, and the same argument explains the larger-mass peak. 
Those estimations also apply to all the following cases. 

\begin{figure}
    \centering
    \includegraphics[width = 0.5\linewidth]{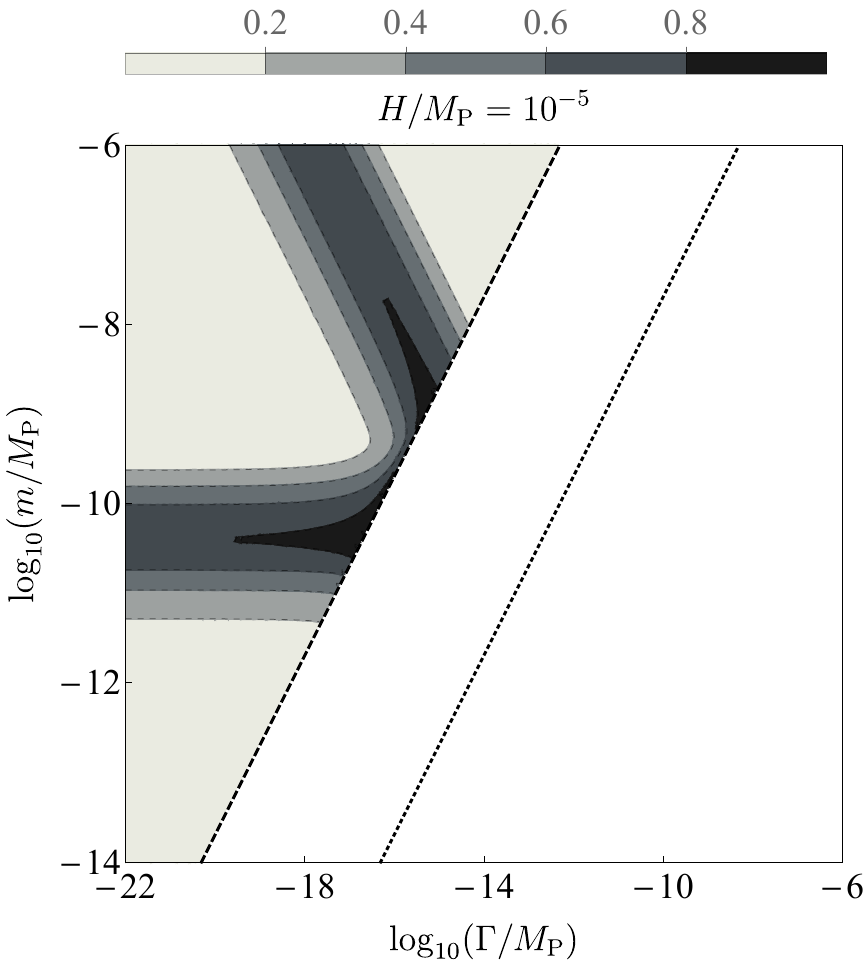}
    \caption{
        The cumulative distribution (probability to realise a non-negligible amount of the contribution from the curvaton) in the parameter space where $\zeta_{\rm obs} < \zeta_{\rm M}$ is satisfied, for the curvaton in the quadratic potential. 
    }
    \label{fig:pmspace_ho}
\end{figure}

Figure~\ref{fig:pmspace_ho} shows the probability to realise a non-negligible contribution of the curvaton in the parameter space, including the possibility to realise the total amount of the curvature perturbation, \textit{i.e.}~$\mathbb{P} (c \, \zeta_{\rm obs} < \zeta < \zeta_{\rm obs})$ defined in Eq.~(\ref{eq:curvaton_cumul}). 
The inflationary energy scale is fixed to $H / M_{\rm P} = 10^{-5}$ in the figure and in what follows, and implementing a different scale merely shifts the high-probability regions. 
The whitish-grey region corresponds to the parameters that give rise to non-negligible (and observationally consistent) curvature perturbation, comprising more than $10\%$ of the total curvature perturbation including $100\%$, but with a low probability, $\mathbb{P} (c \, \zeta_{\rm obs} < \zeta < \zeta_{\rm obs}) < 0.2 = 20\%$. 
In other words, with the probability higher than $80\%$, the parameters in the whitish-grey region would give rise to a negligible contribution and not be of our interest, or excess the observational amplitude to be ruled out in a probabilistic sense, see Figure~\ref{fig:prst_b}. 
The two dark grey and straight regions, on the other hand, represent a relatively high probability larger than $60\%$ but lower than $80\%$ that realises $c \, \zeta_{\rm obs} < \zeta < \zeta_{\rm obs}$. 
Though it is a small black region, there exist the parameters to contribute $\zeta_{\rm obs}$ in a non-negligible way with more than $80\%$ probability. 
This means that with the high probability the set of the parameters in the model can amount to the observed curvature perturbation. 
Finally, the right panel of Figure~\ref{fig:curvaton_dist_ho} is embedded in Figure~\ref{fig:pmspace_ho}, as the vertical line at $\Gamma / M_{\rm P} = 10^{-19}$. 
It is now clear that the horizontal grey region in Figure~\ref{fig:pmspace_ho} corresponds to the ``$+$'' branch where the curvaton dominates the universe at its decay with $r_{\rm decay} \simeq 1$, while the tilted straight line to the ``$-$'' branch where $r_{\rm dec} < 1$. 
This explains why in the former region the high probability can be realised without the dependence on the decay rate, while the order estimation implies that the latter region is along the tilted straight line roughly given by
\begin{equation}
    \log_{10} \qty( \frac{m}{M_{\rm P}} ) + \log_{10} \qty( \frac{\Gamma}{M_{\rm P}} ) \sim 6 \log_{10} \qty( \frac{H}{M_{\rm P}} ) - 2 \log_{10} \zeta 
    \,\, . 
    \label{eq:curv_rstl}
\end{equation} 
Those two branches merge at the boundary (black dashed line) because the two solutions degenerate where $\zeta \approx \zeta_{\rm M}$ and hence $\mathsf{Y} \ll 1$. 

The curvaton in the potential $v (y) = y^{2} / 2 - (\ell + 1) \ln y$ can also be analysed in a similar way. 
This situation corresponds to the quantum-mechanical system in the radial harmonic oscillator. 
The potential of the curvaton expanded around the global minimum gives the quadratic form, $V (\phi) = 6 \alpha^{2} H^{2} ( \phi - \widetilde{\phi} )^{2} + \mathcal{O} ( ( \phi - \widetilde{\phi} )^{3} )$, which enables us to identify $\alpha$ with the mass of the curvaton as $\alpha = (1 / \sqrt{12}) (m / H)$, while the other model parameter $\ell$ remains free. 
Under this identification, the distribution of the curvature perturbation is given by 
\begin{align}
	\mathbb{P} (\zeta) 
	&= \qty( \frac{ \sqrt{2} \, \alpha }{ 3 \zeta_{\rm M}^{2} } ) 
	\frac{2}{\Gamma ( \ell + 3/2)} 
	\qty( \frac{2 \alpha^{2}}{9 \zeta_{\rm M}^{2}} )^{\ell + 1} 
	\frac{1}{\mathsf{Y}} 
	\left\{ 
		\frac{1}{1 - \mathsf{Y}} \qty( \frac{1 + \mathsf{Y}}{1 - \mathsf{Y}} )^{\ell + 1} 
		\exp \qty[ 
			- \qty( \frac{ \sqrt{2} \, \alpha}{3 \zeta_{\rm M}} )^{2} 
			\frac{1 + \mathsf{Y}}{1 - \mathsf{Y}} 
		] 
	\right. \notag \\ 
	&\quad \qquad\qquad\qquad\qquad\qquad\qquad\quad 
	\left. 
		+ \frac{1}{1 + \mathsf{Y}} \qty( \frac{1 - \mathsf{Y}}{1 + \mathsf{Y}} )^{\ell + 1} 
		\exp \qty[ 
			- \qty( \frac{ \sqrt{2} \, \alpha}{3 \zeta_{\rm M}} )^{2} 
			\frac{1 - \mathsf{Y}}{1 + \mathsf{Y}} 
		] 
	\right\} 
	\,\, . 
	\label{eq:curv_dist_rho}
\end{align}
This expression looks similar to Eq.~(\ref{eq:curv_dist_ho}), but the additional factors that contain $\ell$ arise due to the logarithmic correction term in the potential, $(\ell + 1) \ln y$.  
In the limit $\ell \to -1$, it reduces to the one coming from the pure harmonic oscillator (\ref{eq:curv_dist_ho}). 
Indeed, the distribution (\ref{eq:curv_dist_rho}) comes from the $n = 0$ ground-state wavefunction of the corresponding quantum-mechanical system described by the associated Laguerre polynomial, which is related to the lowest-energy state of the Hermite polynomial, see also the previous footnote. 
The curvaton-field value can neither be zero nor negative by definition, so is the curvature perturbation due to the relation (\ref{eq:ph_curv_rel_inv}). 
The normalisation condition is therefore the same as Eq.~(\ref{eq:curv_nom_ho}), but in this case the additional factor two is absent in Eq.~(\ref{eq:curv_dist_rho}). 

\begin{figure}
    \begin{subfigure}[b]{0.49\textwidth}
        \centering
        \includegraphics[width = 0.995\linewidth]{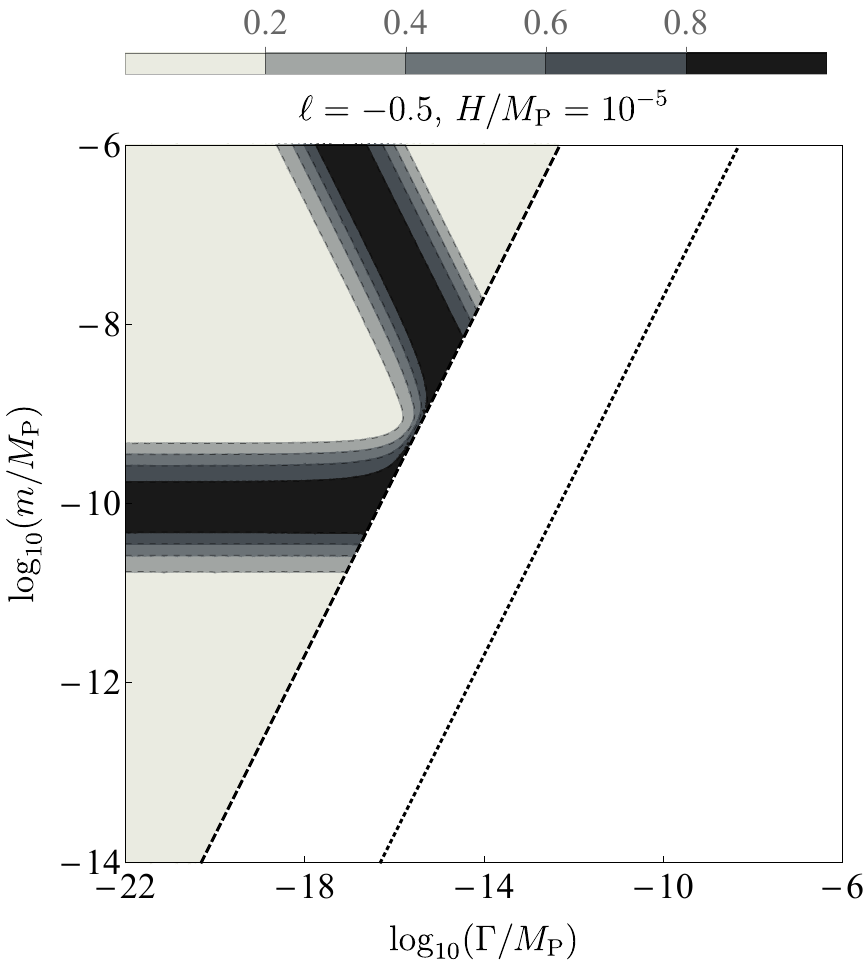}
    \end{subfigure}
    \begin{subfigure}[b]{0.49\textwidth}
        \centering
        \includegraphics[width = 0.995\linewidth]{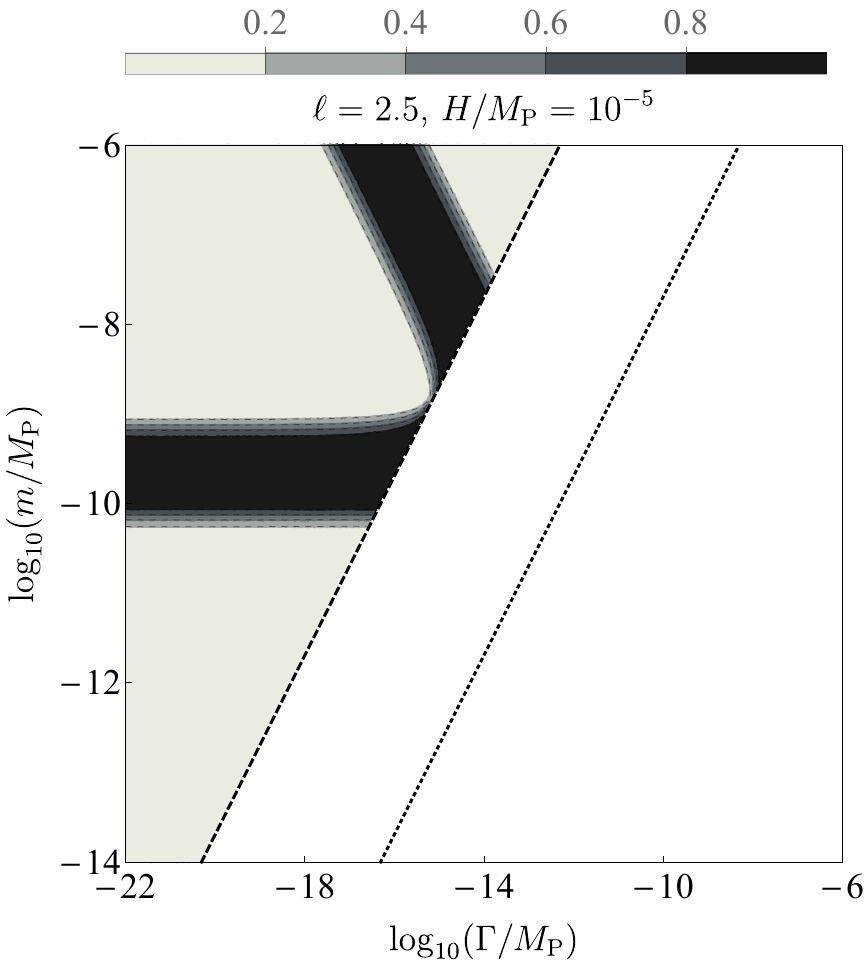}
    \end{subfigure}
    \caption{
		The cumulative distribution for the curvaton in the quadratic-logarithmic potential for $\ell = - 0.5$ (\textit{left}) and $\ell = 2.5$ (\textit{right}).  
	}
    \label{fig:curvaton_rho_pmspace}
\end{figure}

Figure~\ref{fig:curvaton_rho_pmspace} shows the cumulative distribution for $\ell = - 0.5$ and $2.5$ in the left and right panels respectively. 
The integration domain of $\zeta$ to define the probability is also the same as Figure~\ref{fig:pmspace_ho}, realising at least $10\%$ contribution from the curvaton that can even amount to $100\%$. 
The region with high probability in the parameter space (black region) is similar to the previous case in the sense that it is limited in the two specific lines. 
What is different from the previous case is a broader black region can be found, meaning that there exists more combination of the set of the parameters $(\Gamma, \, m)$ that gives rise to a non-negligible fraction of the total $\zeta_{\rm obs}$ with a probability larger than $80\%$. 
The additional parameter $\ell$ controls the width of those high-probability regions, and as long as $\ell > 0$, larger $\ell$ gives broader high-probability region. 
On the other hand, the high-probability region shrinks and converges to reproduce Figure~\ref{fig:pmspace_ho} as $\ell$ approaches to $-1$. 

\subsubsection{Curvaton in trigonometric potential}
\label{subsec:curvaton_tri}

Next, let us study the parameter-space structure of the curvaton model with a class of trigonometric-type potentials. 
Since, as will be seen at the end of this section, in the symmetric ($t = 0$) limit the difference between the trigonometric-Scarf and the trigonometric-Rosen--Morse cases vanishes up to the normalisation factor, the detailed analyses in the following are performed for the trigonometric-Scarf case.

The trigonometric Scarf potential for the curvaton (in the sense that the corresponding quantum-mechanical system is described by the trigonometric Scarf potential) also yields the quadratic form around the global minimum\footnote{
    The expansion of $V (\phi)$ around the bottom up to the quartic term is $V (\phi) = 3 \alpha^{2} s H^{2} \phi^{2} + 4 \pi^{2} \alpha^{2} s \phi^{4} + \mathcal{O} (\phi^{6})$ when $t = 0$. 
}, $V (\phi) = 3 \alpha^{2} s H^{2} (\phi - \widetilde{\phi})^{2}  + \mathcal{O} ( (\phi - \widetilde{\phi} )^{3} )$, which leads us to identify the model parameter with $\alpha = (1 / \sqrt{6 s} \,) (m/H)$. 
Note that this identification does not depend on the asymmetric parameter $t$ in the potential even when $t \neq 0$, while the location of the global minimum $\widetilde{\phi}$ does. 
With this identification, the distribution of the curvature perturbation reads 
\begin{align}
	\mathbb{P} (\zeta)
	&= \frac{2 s}{2^{2s}} \frac{ 2 \Gamma (2s) }{ \Gamma (s - t + 1/2) \Gamma (s + t + 1/2) } \qty( \frac{ \sqrt{2} \, \alpha }{ 3 \zeta_{\rm M}^{2}} ) 
	\notag \\ 
	&\quad 
	\times 
	\frac{1}{\mathsf{Y}} 
	\left\{ 
		\frac{1}{1 - \mathsf{Y}} 
		\qty[ 
			1 - \sin \qty( \frac{ \sqrt{2} \, \alpha }{ 3 \zeta_{\rm M} } \sqrt{ \frac{1 + \mathsf{Y}}{1 - \mathsf{Y}} } \, ) 
		]^{s-t} 
		\qty[ 
			1 + \sin \qty( \frac{ \sqrt{2} \, \alpha }{ 3 \zeta_{\rm M} } \sqrt{ \frac{1 + \mathsf{Y}}{1 - \mathsf{Y}} } \, ) 
		]^{s+t} 
	\right. 
	\notag \\ 
	&\quad \quad \ \,\,
	+ 
	\left. 
		\frac{1}{1 + \mathsf{Y}} 
		\qty[ 
			1 - \sin \qty( \frac{ \sqrt{2} \, \alpha }{ 3 \zeta_{\rm M} } \sqrt{ \frac{1 - \mathsf{Y}}{1 + \mathsf{Y}} } \, ) 
		]^{s-t} 
		\qty[ 
			1 + \sin \qty( \frac{ \sqrt{2} \, \alpha }{ 3 \zeta_{\rm M} } \sqrt{ \frac{1 - \mathsf{Y}}{1 + \mathsf{Y}} } \, ) 
		]^{s+t} 
	\right\} 
\end{align}
Hereafter, the symmetric situation (\textit{i.e.}~$t = 0$) is considered for simplicity. 
The distribution then reduces to 
\begin{equation}
    \mathbb{P} (\zeta)
	= \frac{2 s}{2^{2s}} \frac{ 2 \Gamma (2s) }{ \qty[ \Gamma (s + 1/2) ]^{2} } \qty( \frac{ \sqrt{2} \, \alpha }{ 3 \zeta_{\rm M}^{2}} ) 
    \frac{1}{\mathsf{Y}} 
    \qty[ 
        \frac{1}{1 - \mathsf{Y}} 
        \cos^{2 s} \qty( \frac{ \sqrt{2} \, \alpha }{3 \zeta_{\rm M}} \sqrt{ \frac{1 + \mathsf{Y}}{1 - \mathsf{Y}} } \, ) 
        + \frac{1}{1 + \mathsf{Y}} 
        \cos^{2 s} \qty( \frac{ \sqrt{2} \, \alpha }{3 \zeta_{\rm M}} \sqrt{ \frac{1 - \mathsf{Y}}{1 + \mathsf{Y}} } \, ) 
    ] 
    \label{eq:curv_dist_tsc}
\end{equation}
The additional factor two again comes from the correct normalisation requirement, which will be discussed from now. 

\begin{figure}
    \centering
    \includegraphics[width = 0.5\linewidth]{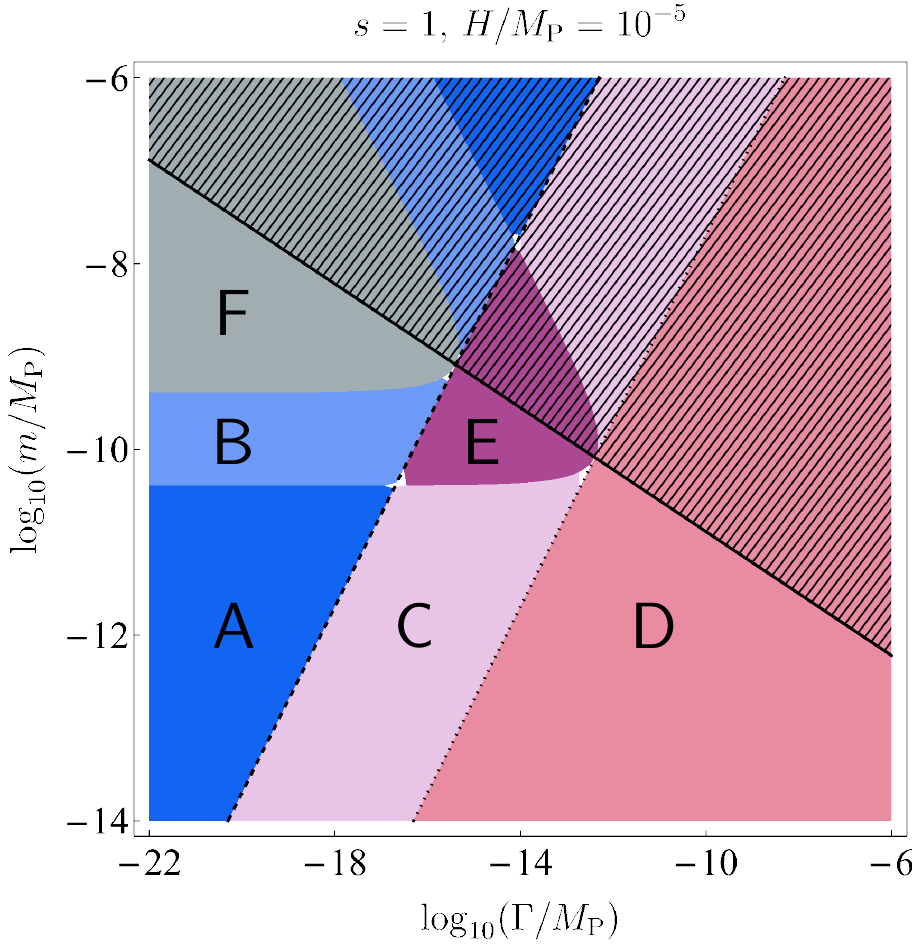}
    \caption{
        The parameter-space of the curvaton model with the trigonometric potentials (\textit{i.e.}~trigonometric Scarf and Rosen--Morse potentials) is divided into six regions, labelled with $\mathsf{A}$, $\dots$, $\mathsf{F}$, see also Table~\ref{tab:cbc}. 
        For the parameters in the region with the solid diagonal lines, the correspondence between $\phi$ and $\zeta$ in Eq.~(\ref{eq:ph_curv_rel}) does not exist, hence the region is inadequate. 
        The integration over $\zeta$ differs in each region, as summarised in Table~\ref{tab:cbc}. 
    }
    \label{fig:prst_tri}
\end{figure}

The parameter-space structure is rather complicated in the present case due to the restriction of the available field-value range. 
First, let us consider the normalisation of $\mathbb{P} (\zeta)$. 
Expressing the relation (\ref{eq:ph_curv_rel_inv}) in terms of $y = \alpha \sqrt{8 \pi^{2} / H^{2}} \, \phi$, one gets 
\begin{equation}
    y_{\pm} 
    = \frac{\sqrt{2} \, \alpha}{3 \zeta} \qty[ 
        1 \pm \sqrt{ 1 - \qty( \frac{\zeta}{\zeta_{\rm M}} )^{2} } 
        \, 
    ] 
    \,\, . 
\end{equation}
It is required that both the $+$ and $-$ branches come inside the region $- \pi/2 < y_{\pm} < \pi/2$. 
This restriction was absent in the harmonic potentials considered before, and makes the parameter space more diverse as will be seen below. 
Upon normalising $\mathbb{P} (\zeta)$ in the domain corresponding to $0 < y < \pi / 2$, only the region that satisfies 
\begin{equation}
    x \coloneqq \frac{2}{\pi} \qty( \frac{ \sqrt{2} \, \alpha }{ 3 \zeta_{\rm M} } ) < 1 
    \label{eq:curv_tsc_avreg}
\end{equation}
is allowed. 
The boundary is given by the straight line, 
\begin{equation}
    \log_{10} \qty( \frac{m}{M_{\rm P}} ) 
    = - \frac{1}{3} \log_{10} \qty( \frac{\Gamma}{M_{\rm P}} ) + \frac{8}{3} \log_{10} \qty( \frac{H}{M_{\rm P}} ) - \frac{4}{3} \log_{10} \qty( \frac{8}{\sqrt{3 s}} ) 
    \,\, , 
\end{equation}
under which is allowed, while the inadequate region is shown in Figure~\ref{fig:prst_tri} by the diagonal and negative-tilted solid line. 
Under the condition (\ref{eq:curv_tsc_avreg}), for a set of the parameters $(\Gamma, \, m)$ to give the reasonable value of $y$, they must lie inside the region, 
\begin{equation}
    \frac{2 x}{1 + x^{2}} < \frac{\zeta}{\zeta_{\rm M}} < 1 \,\, , 
\end{equation}
where $0 < 2 x / (1 + x^{2}) < 1$ as long as the condition (\ref{eq:curv_tsc_avreg}) is satisfied. 
Given those restrictions, the adequate normalisation condition reads 
\begin{equation}
    \int_{ [2 x/(1+x^{2})] \zeta_{\rm M} }^{\zeta_{\rm M} } \dd \zeta \, \mathbb{P} (\zeta) 
    = 1 
    \,\, . 
\end{equation}

Next, let us take into account the interplay of $\zeta_{\rm obs}$ with the domain to be normalised. 
Under the condition $x < 1$, there are four quantities that can be ordered in six different ways, the two are of the normalisation domain $[ 2 x / (1 + x^{2}) ] \zeta_{\rm M} < \zeta_{\rm M}$ and the other two are of the curvature perturbation $c \, \zeta_{\rm obs} < \zeta_{\rm obs}$, where $0 < c < 1$. 
The interplay amongst those quantities divides the parameter space into six regions as shown in Figure~\ref{fig:prst_tri}. 
Each order in each region determines the cumulative distribution (\ref{eq:curvaton_cumul}) in the overlapped domain defined by $[ 2 x / (1 + x^{2}) ] \zeta_{\rm M} < \zeta < \zeta_{\rm M}$ and $c \, \zeta_{\rm obs} < \zeta < \zeta_{\rm obs}$. 
The six situations are as follows. 
As in the previous cases, the fraction $0 < c < 1$ is fixed to be $c = 0.1$. 
\begin{table}
    \renewcommand{\arraystretch}{2.0}
	\centering 
	\caption{
		The categorisation of the six parameter-space regions labelled as $\mathsf{A}$ to $\mathsf{F}$ in Figure.~\ref{fig:prst_tri}, domain to be integrated, and how much the curvaton can contribute to the observed amount of the curvature perturbation. 
		\\ 
	}
	\begin{tabular}{|c|c|c|c|}
		\hline 
		\textbf{Region} & \textbf{Condition} & \textbf{Integration} & \textbf{Curvaton Contribution} \\ 
		\hline \hline 
		$\mathsf{A}$ & $\displaystyle \frac{2 x}{1 + x^{2}} < c \, \frac{\zeta_{\rm obs}}{\zeta_{\rm M}} < \frac{\zeta_{\rm{obs}}}{\zeta_{\rm M}} < 1$ & $c \, \zeta_{\rm obs} < \zeta < \zeta_{\rm obs}$ & At least $10\%$ and amount to $100\%$ \\ 
		\hline 
        $\mathsf{B}$ & $\displaystyle c \, \frac{\zeta_{\rm obs}}{\zeta_{\rm M}} < \frac{2 x}{1 + x^{2}} < \frac{\zeta_{\rm obs}}{\zeta_{\rm M}} < 1$ & $\displaystyle \frac{2 x}{1 + x^{2}} \zeta_{\rm M} < \zeta < \zeta_{\rm obs}$ & More than $10\%$ and amount to $100\%$ \\ 
        \hline 
        $\mathsf{C}$ & $\displaystyle \frac{2 x}{1 + x^{2}} < c \, \frac{\zeta_{\rm obs}}{\zeta_{\rm M}} < 1 < \frac{\zeta_{\rm obs}}{\zeta_{\rm M}}$ & $\displaystyle c \, \zeta_{\rm obs} < \zeta < \zeta_{\rm M}$ & At least $10\%$ but less than $100\%$ \\
        \hline
        $\mathsf{D}$ & $\displaystyle \frac{2 x}{1 + x^{2}} < 1 < c \, \frac{\zeta_{\rm obs}}{\zeta_{\rm M}} < \frac{\zeta_{\rm obs}}{\zeta_{\rm M}}$ & $\displaystyle \frac{2 x}{1 + x^{2}} \zeta_{\rm M} < \zeta < \zeta_{\rm M}$ & Less than $10\%$ \\
        \hline
        $\mathsf{E}$ & $\displaystyle c \, \frac{\zeta_{\rm obs}}{\zeta_{\rm M}} < \frac{2 x}{1 + x^{2}} < 1 < \frac{\zeta_{\rm obs}}{\zeta_{\rm M}}$ & $\displaystyle \frac{2 x}{1 + x^{2}} \zeta_{\rm M} < \zeta < \zeta_{\rm M}$ & More than $10\%$ but less than $100\%$ \\
        \hline
        $\mathsf{F}$ & $\displaystyle c \, \frac{\zeta_{\rm obs}}{\zeta_{\rm M}} < \frac{\zeta_{\rm obs}}{\zeta_{\rm M}} < \frac{2 x}{1 + x^{2}} < 1$ & --- & More than $100\%$ \\
        \hline
	\end{tabular}
	\label{tab:cbc}
\end{table}

\begin{enumerate}
    \item[$\mathsf{A}$.]
        $[ 2 x / (1 + x^{2}) ] \zeta_{\rm M} < c \, \zeta_{\rm obs} < \zeta_{\rm obs} < \zeta_{\rm M}$. 
        Both the $10\%$ and $100\%$ of the curvature perturbation come inside the domain to be normalised in this case. 
        The integration to define the cumulative distribution is therefore done in the domain $c \, \zeta_{\rm obs} < \zeta < \zeta_{\rm obs}$. 
        For each set of the parameters in the region $\mathsf{A}$, this gives the probability that the total amount of the curvature perturbation consists of at least $10\%$ origin from the curvaton, including the possibility of constituting the total $\zeta_{\rm obs}$. 
        
    \item[$\mathsf{B}$.] 
        $c \, \zeta_{\rm obs} < [ 2 x / (1 + x^{2}) ] \zeta_{\rm M} < \zeta_{\rm obs} < \zeta_{\rm M}$. 
        In this case, the $10\%$ value of the total curvature perturbation does not come inside the integration domain to define the cumulative distribution $\mathbb{P} ([ 2 x / (1 + x^{2}) ] \zeta_{\rm M} < \zeta < \zeta_{\rm obs})$, but instead the $100\%$ value does probabilistically ensuring more than $10\%$ contribution. 
        This means that a parameter set inside the region $\mathsf{B}$ necessarily produces a non-negligible contribution to the total $\zeta_{\rm obs}$ with a specific probability depending on $(\Gamma, \, m)$. 
        
    \item[$\mathsf{C}$.]
        $[ 2 x / (1 + x^{2}) ] \zeta_{\rm M} < c \, \zeta_{\rm obs} < \zeta_{\rm M} < \zeta_{\rm obs}$. 
        The total amount of the observed curvature perturbation comes outside the domain to be integrated, $c \, \zeta_{\rm obs} < \zeta < \zeta_{\rm M}$, which implies that another source than that from the curvaton is needed to achieve $\zeta_{\rm obs}$ although $10\%$ of the total can be realised.

    \item[$\mathsf{D}$.]
        $[ 2 x / (1 + x^{2}) ] \zeta_{\rm M} < \zeta_{\rm M} < c \, \zeta_{\rm obs} < \zeta_{\rm obs}$. 
        This order of the quantities corresponds to the parameters in the region $\mathsf{C}$ in Figure~\ref{fig:prst}, and is not of interest. 
        Those parameter sets can at most negligibly contribute to the total curvature perturbation and, hence will not be of the least interest. 
        
    \item[$\mathsf{E}$.]
        $c \, \zeta_{\rm obs} < [ 2 x / (1 + x^{2}) ] \zeta_{\rm M} < \zeta_{\rm M} < \zeta_{\rm obs}$. 
        In this case, the distribution is integrated in its whole domain to give unity (since it matches the domain to be normalised), implying that the total curvature perturbation cannot originate from the curvaton. 
        Different from $\mathsf{D}$, however, a non-negligible contribution that means more than $10\%$ of the total $\zeta_{\rm obs}$ from the curvaton is ensured. 
        
    \item[$\mathsf{F}$.]
        $c \, \zeta_{\rm obs} < \zeta_{\rm obs} < [ 2 x / (1 + x^{2}) ] \zeta_{\rm M} < \zeta_{\rm M}$. 
        For a set of the parameters $(\Gamma, \, m)$ in this region, the integral $\mathbb{P} ([ 2 x / (1 + x^{2}) ] \zeta_{\rm M} < \zeta < \zeta_{\rm M})$ is again unity, but now implying the overproduction of the curvature perturbation and hence necessarily contradicting to the observation. 
\end{enumerate}

\begin{figure}
    \begin{subfigure}[b]{0.49\textwidth}
        \centering
        \includegraphics[width = 0.995\linewidth]{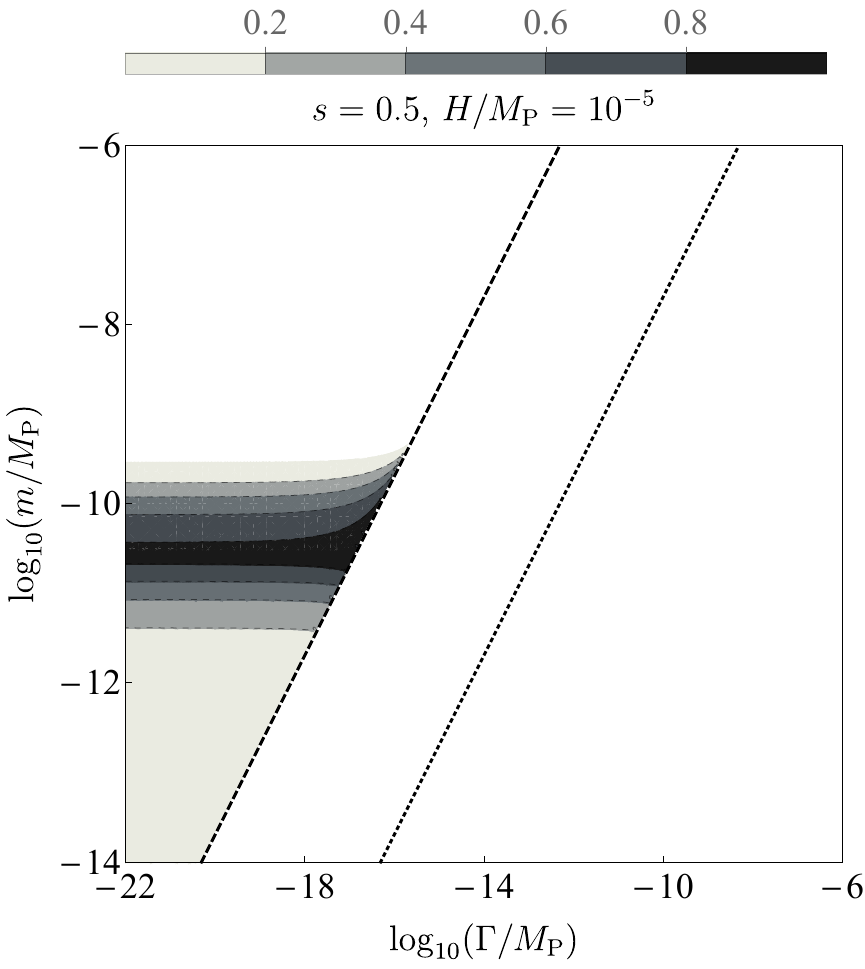}
    \end{subfigure}
    \begin{subfigure}[b]{0.49\textwidth}
        \centering
        \includegraphics[width = 0.995\linewidth]{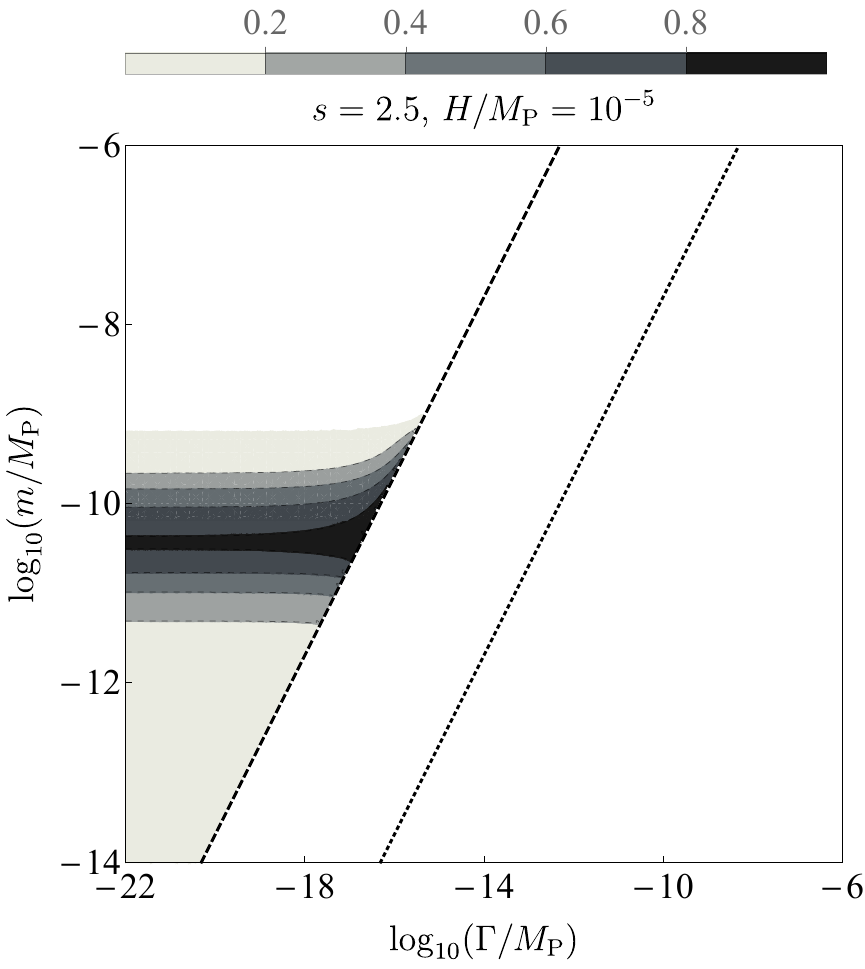}
    \end{subfigure}
    \caption{
		The cumulative distribution for the curvaton in the trigonometric Scarf potential for $s = 0.5$ (\textit{left}) and $s = 2.5$ (\textit{right}).  
	}
    \label{fig:curvaton_tsc_pmspace}
\end{figure}

Figure~\ref{fig:prst_tri} and Table~\ref{tab:cbc} summarise these six regions, labelled by $\mathsf{A}, \, \dots, \, \mathsf{F}$. 
Based on this dividing of the whole parameter space, Figure~\ref{fig:curvaton_tsc_pmspace} shows the cumulative distribution for $s = 0.5$ and $s = 2.5$ in the left and right panels respectively. 
The plot is made for the parameters in the region $\mathsf{A}$ and $\mathsf{B}$ in Figure~\ref{fig:prst_tri}. 
One branch corresponding to $r_{\rm dec} < 1$ vanishes, which was present in the harmonic-potential cases. 
It can be seen that for a larger $s$, the width of the mass that gives a high probability larger than $80\%$ gets narrower. 
Indeed, in the limit $s \to \infty$, the shown parameter region reduces to that of the curvaton in the quadratic potential shown in Figure~\ref{fig:pmspace_ho}, see also Appendix~\ref{appx:appx2}. 
The specific mass that gives the high-probability region regardless of the decay rate, can be known by the same order-of-magnitude estimation as that done for the harmonic oscillator case. 
The standard deviation squared (variance) again reads $\expval{ (\phi / H)^{2} }_{\infty} \sim (H / m)^{2}$, which is enough to estimate the order of the preferred mass. 
In a more rigorous way, one can carry out the integral (\ref{eq:exs_stdist_sc1_c}) when $t = 0$ to derive the quantitative difference of the trigonometric-Scarf case from the harmonic-oscillator case (see Appendix~\ref{appx:appx2}), 
\begin{equation}
    \expval{ \qty( \frac{\phi}{H} )^{2} }_{\infty} 
    = \int \dd \phi \, \qty( \frac{\phi}{H} )^{2} f_{\infty} (\phi) 
    = \frac{3}{8 \pi^{2}} \qty[ s \sum_{k=1}^{\infty} \frac{1}{(s + k)^{2}} ] \qty( \frac{H}{m} )^{2} 
    \,\, . 
    \label{eq:tris_quadlim}
\end{equation}
Given Eqs.~(\ref{eq:exs_mv}), the function with respect to $s$ in the squared bracket measures the deviation of the variance from that of the curvaton in the pure quadratic potential. 
It is a monotonically increasing function with respect to $s$, and, in particular, goes to unity when $s \to \infty$. 
In this limit, the variance of the trigonometric Scarf and harmonic oscillator cases coincide with each other. 
This is because the trigonometric Scarf potential gets steeper as $s > 0$ becomes large, so the curvaton tends to be confined near the bottom of the potential, where it is approximated by the quadratic function. 
Therefore, the curvaton cannot tell whether it is confined in the quadratic or trigonometric Scarf potential in the $s \to \infty$ limit. 
A cumbersome derivation of Eq.~(\ref{eq:tris_quadlim}) as well as $\expval{ \phi^{4} }_{\infty}$ can be found in Appendix~\ref{appx:appx2}. 

Before closing this section, let us give the distribution function of the curvature perturbation originating from the curvaton in the potential $v (y) = - (t/s) y - s \ln \cos y$ (corresponding to the trigonometric Rosen--Morse potential in quantum mechanics) for completeness. 
From the same argument in the three previous cases, the distribution reads 
\begin{align}
	\mathbb{P} (\zeta) 
	&= \frac{2^{2 s}}{\pi} 
	\frac{ \Gamma (s + 1 - i t/s) \Gamma (s + 1 + i t/s) }{s \Gamma (2s)} 
	\qty( \frac{ \sqrt{2} \, \alpha }{ 3 \zeta_{\rm M}^{2} } ) 
	\notag \\ 
	&\quad 
	\times 
	\frac{1}{\mathsf{Y}} 
	\left\{ 
		\frac{1}{1 - \mathsf{Y}} 
		\cos^{2s} \qty( \frac{\sqrt{2} \, \alpha}{3 \zeta_{\rm M}} \sqrt{ \frac{ 1 + \mathsf{Y} }{ 1 - \mathsf{Y} } } \, ) 
		\exp \qty[ 
			\qty( \frac{\sqrt{2} \, \alpha}{3 \zeta_{\rm M}} ) 
			\frac{2 t}{s} 
			\sqrt{ \frac{ 1 + \mathsf{Y} }{ 1 - \mathsf{Y} } } 
		] 
	\right. 
	\notag \\ 
	&\quad \quad \ \,\,
	+ 
	\left. 
		\frac{1}{1 + \mathsf{Y}} 
		\cos^{2s} \qty( \frac{\sqrt{2} \, \alpha}{3 \zeta_{\rm M}} \sqrt{ \frac{ 1 - \mathsf{Y} }{ 1 + \mathsf{Y} } } \, ) 
		\exp \qty[ 
			\qty( \frac{\sqrt{2} \, \alpha}{3 \zeta_{\rm M}} ) 
			\frac{2 t}{s} 
			\sqrt{ \frac{ 1 - \mathsf{Y} }{ 1 + \mathsf{Y} } } 
		] 
	\right\} 
    \,\, . 
\end{align}
In the $t = 0$ (symmetric) limit, the two exponentials vanish and it reduces to 
\begin{equation}
    \mathbb{P} (\zeta) 
    = \frac{2^{2 s}}{\pi} 
	\frac{ \qty[ \Gamma (s + 1) ]^{2}}{s \Gamma (2s)} 
	\qty( \frac{ \sqrt{2} \, \alpha }{ 3 \zeta_{\rm M}^{2} } ) 
    \frac{1}{\mathsf{Y}} 
    \qty[ 
        \frac{1}{1 - \mathsf{Y}} 
		\cos^{2s} \qty( \frac{\sqrt{2} \, \alpha}{3 \zeta_{\rm M}} \sqrt{ \frac{ 1 + \mathsf{Y} }{ 1 - \mathsf{Y} } } \, ) 
        + 
        \frac{1}{1 + \mathsf{Y}} 
		\cos^{2s} \qty( \frac{\sqrt{2} \, \alpha}{3 \zeta_{\rm M}} \sqrt{ \frac{ 1 - \mathsf{Y} }{ 1 + \mathsf{Y} } } \, ) 
    ] \,\, . 
    \label{eq:curv_dist_trm}
\end{equation}
The parameter $\alpha$ and $s$ can be related to the mass according to $\alpha = (1 / \sqrt{6 s} \, ) (m / H)$, which is the same as in the trigonometric Scarf case. 
One notices that Eq.~(\ref{eq:curv_dist_trm}) is almost the same as the distribution (\ref{eq:curv_dist_tsc}) for the trigonometric Scarf potential except the prefactor, which depends only on the model parameter $s$ and does include neither the decay rate nor the mass of the curvaton. 
However, once the asymmetric parameter $t$ is taken into account, non-trivial deformation of the parameter-space structure compared to the previous case would be expected. 
This investigation is left for future work. 

\section{Conclusion}
\label{sec:concl}

The origin of all the present cosmological structures can be traced back to the quantum fluctuations generated during inflation. 
Classicalisation of those fluctuations to lose their quantumness, the notion of horizon crossing, motivates us to construct an effective field-theoretical treatment for large-scale field configurations~\cite{Starobinsky:1986fx}, so-called the stochastic formalism of inflation. 
Due to the correspondence between the diffusion and Schr\"{o}dinger equations, all the possible exact solutions in stochastic inflation can exhaustively be listed, in the sense that the corresponding wavefunctions are expressed in terms of classical orthogonal polynomials. 
A class of those exact solutions are explicitly and concretely presented in~\cite{Honda:2024evc}. 

The present paper demonstrates an application of those exact solutions to the curvaton paradigm \cite{Lyth:2001nq, Moroi:2001ct, Enqvist:2001zp, Lyth:2002my}, in which the curvaton can explain a part of or all the observed curvature perturbation, relieving the inflaton from the responsibility of generating it. 
Assuming that the inflationary universe is settled down to its stationary state to make our analyses as simple as possible, a test field is identified with the curvaton in section~\ref{sec:curvaton}. 
By doing so, from the functional relation between the curvaton and the curvature perturbation, and from the analytical stationary distributions of a test field in the stochastic formalism of inflation reviewed in section~\ref{subsec:exactsol}, the corresponding distributions of the curvature perturbation are presented. 
Normalising it and defining the probability to realise a specific amount of the curvature perturbation from the curvaton, the non-trivial structure of the parameter space is analysed in section~\ref{subsec:curvaton_harmonic} and \ref{subsec:curvaton_tri}. 
For the harmonic-type potentials, the parameter space is divided into the three regions, based on how much the total curvature perturbation can originate from the curvaton. 
One of those three regions is focussed on because the decay rate and the mass of the curvaton in a subset of the region can realise the total amount of the observed curvature perturbation, as well as ensure a non-negligible contribution of the curvaton in a probabilistic sense. 
On the other hand, the trigonometric-type potentials make the parameter-space structure more complicated due to the restriction of the possible field values, dividing the parameter space into six regions. 
Nevertheless, there exists a set of the parameters that can give rise to the total curvature perturbation with a high probability as well. 
For the potentials other than the pure harmonic oscillator, the additional model parameter (such as $\ell$ in the radial-harmonic-oscillator case and $s$ in the trigonometric potentials) can affect and extend or curtail a part of the region that can realise the non-negligible contribution of the curvaton with a high probability. 

This paper concludes by mentioning several possible future directions. 
First, the relation between the curvaton field and the curvature perturbation, given by Eq.~(\ref{eq:ph_curv_rel_inv}), is exact only when the potential is quadratic (and at leading order of the fluctuations). 
For a non-quadratic potential, there may be possible correction terms that can affect the observables such as non-Gaussianities, see \textit{e.g.}~Ref.~\cite{Kawasaki:2011pd}. 
Since the potentials considered in the present paper include the non-quadratic ones (although around the bottom all of the potentials can be approximated by the quadratic function), it would be interesting to refine the quadratic approximation (\ref{eq:ph_curv_rel_inv}) and study the possible modifications of the parameter space. 
Deriving the distribution of the curvature perturbation and studying the non-trivial structure of the parameter space with the effect coming from non-Gaussianities in addition to the non-quadratic correction being incorporated would also be interesting, as was partially done in~\cite{Lerner_2014}. 
Second, although the stationary distribution will extensively be applied for our first demonstration of the exact solutions in stochastic inflation, the time evolution of the distribution and parameter space can be kept track of. 
In such cases, the initial condition of the curvaton field can affect the outcomes. 
This is in contrast to the present analyses where the initial value of the curvaton field is irrelevant, and hence would be worth investigating. 

\acknowledgments

The author is grateful to Ryusuke Jinno, Masahiro Kawasaki, and Tomo Takahashi for insightful discussion and for fruitful comments on the manuscript, and was financially supported by the Sasakawa Scientific Research Grant from The Japan Science Society (JSS) and by the Japan Society for the Promotion of Science (JSPS) KAKENHI Grant Number JP24K22877. 

\appendix

\section{Other regions in parameter space}
\label{appx:appx1}

While in the main text the analyses of the parameter space are restricted to region $\mathsf{A}$, a similar discussion can be made for the other regions (as well as for other choices of the domain to be integrated). 
One example amongst them focussing on region $\mathsf{B}$ in addition to $\mathsf{A}$ when the curvaton in the pure quadratic potential, $v (y) = y^{2} / 2$, is discussed here. 

\begin{figure}
    \centering
    \includegraphics[width = 0.5\linewidth]{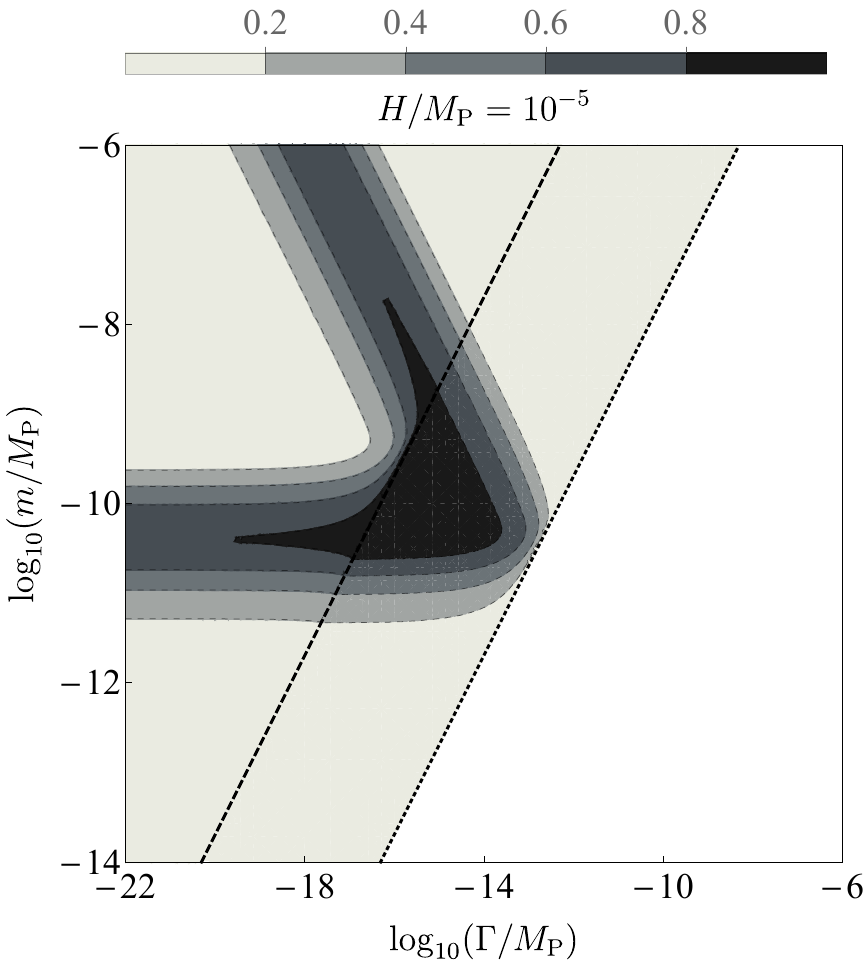}
    \caption{
        The cumulative distribution (probability to realise a non-negligible amount of the contribution from the curvaton) in the region $\mathsf{A}$ and $\mathsf{B}$, for the curvaton in the pure quadratic potential given by $v (y) = y^{2} / 2$. 
    }
    \label{fig:pmspace_ho-2}
\end{figure}

The cumulative distribution with $\zeta_{1} = c \, \zeta_{\rm obs}$ and $\zeta_{2} = \zeta_{\rm obs}$, \textit{i.e.}~$\displaystyle \mathbb{P} (\zeta_{1} < \zeta < \zeta_{2}) = \int_{c \, \zeta_{\rm obs}}^{\zeta_{\rm obs}} \dd \zeta \, \mathbb{P} (\zeta)$, is plotted in region $\mathsf{A}$ in Figure~\ref{fig:pmspace_ho-2}, which is exactly the same as Figure~\ref{fig:pmspace_ho}. 
For the parameters in the region $\mathsf{B}$, on the other hand, the total curvature perturbation cannot be the upper limit of the integration since $\zeta_{\rm M} < \zeta_{\rm obs}$, see Figure~\ref{fig:prst_b}. 
Instead, the cumulative distribution with $\zeta_{1} = c \, \zeta_{\rm obs}$ and $\zeta_{2} = \zeta_{\rm M}$, \textit{i.e.}~$\displaystyle \mathbb{P} (\zeta_{1} < \zeta < \zeta_{2}) = \int_{c \, \zeta_{\rm obs}}^{\zeta_{\rm M}} \dd \zeta \, \mathbb{P} (\zeta)$, is displayed in the region $\mathsf{B}$. 
The upper limit of the integration is different between the two regions, but it continuously changes from $\zeta_{\rm obs}$ to $\zeta_{\rm M}$ as the parameters $(\Gamma, \, m)$ crosses the dividing line between the region $\mathsf{A}$ and $\mathsf{B}$ (the dashed straight line), and $\zeta_{\rm M} = \zeta_{\rm obs}$ holds on the boundary in particular. 
This is why the coloured regions are continuous on both sides of the line. 
The figure indicates that not only in the region $\mathsf{A}$ but also $\mathsf{B}$ there exist parameters that can give rise to a non-negligible contribution to the observed curvature perturbation (especially the black region). 
Meanwhile, it should be noted that the meaning of the probability differs between the regions. 
Specifically, the integral $\mathbb{P} (c \, \zeta_{\rm obs} < \zeta < \zeta_{\rm obs})$ in the region $\mathsf{A}$ measures the probability that the curvaton can realise at least $10\%$, including the total, of the observed curvature perturbation, while the integral $\mathbb{P} (c \, \zeta_{\rm obs} < \zeta < \zeta_{\rm M})$ in the region $\mathsf{B}$ does the probability but another source for $\zeta_{\rm obs}$ is also required to explain the observed amplitude of the curvature perturbation. 
This is the reason that the analyses are restricted to region $\mathsf{A}$ in the main text, although a similar discussion can be made for the other regions that can give rise to a part of the total curvature perturbation. 

In a similar manner, the integral $\mathbb{P} (0 < \zeta < c \, \zeta_{\rm obs})$ and $\mathbb{P} (\zeta_{\rm obs} < \zeta < \zeta_{\rm M})$ for instance, can also be displayed in region $\mathsf{A}$ instead of $\mathbb{P} (c \, \zeta_{\rm obs} < \zeta < \zeta_{\rm obs})$, see Figure~\ref{fig:prst_b}. 
The former measures the probability that the curvature perturbation generated from the curvaton cannot amount to less than even $10\%$ of the total, while the latter does the probability that it exceeds the observed amplitude. 

\section{Statistical moments in trigonometric-Scarf case}
\label{appx:appx2}

The statistical moments $\expval{ \phi^{n} }_{\infty}$ in the case of the trigonometric Scarf and the relation between the pure harmonic oscillator and the trigonometric Scarf are considered here. 
To obtain the analytical expressions of the statistical moments when $t = 0$, it is necessary to compute the definite integral in Eq.~(\ref{eq:exs_stdist_sc1_c}), 
\begin{equation}
    \int_{- \pi / 2}^{\pi / 2} \dd y \, y^{n} \cos^{2 s} y 
    \,\, . 
\end{equation}

\subsection{General formula}

For any non-negative integer $n$ and any positive number $s$, the following indefinite integral can be conjectured, 
\begin{equation}
	\int \dd y \, y^{n} \cos^{2 s} y 
	= \frac{e^{- 2 i s y}}{2^{2 s} s^{n+1}} 
	\sum_{p=1}^{n+1} 
	\frac{(- i)^{f_{4} (p+2)}}{2^{p}} 
	\frac{\Gamma (n+1)}{\Gamma (n - p + 2)} 
    (s y)^{n - p + 1} 
    \cdot 
	{}_{p+1} F_{p} 
    \qty(
        \begin{matrix}
            -2 s, \, -s, \, \dots, \, -s 
            \\ 
            1-s, \, \dots, \, 1-s 
        \end{matrix}
        ~ \middle| ~ - e^{2 i y} 
    ) 
	\,\, . 
	\label{eq:conj_master}
\end{equation}
In Eq.~(\ref{eq:conj_master}), $f_{4} (a) = 0$, $1$, $2$, or $3$ is the remainder when $a$ is divided by four. 
The generalised hypergeometric function is defined by 
\begin{equation}
    {}_{r} F_{s} 
    \qty(
        ~ 
        \begin{matrix}
            a_{1}, \, \dots, \, a_{r} 
            \\ 
            b_{1}, \, \dots, \, b_{s} 
        \end{matrix} 
        ~ \middle| ~ z 
    ) 
    \coloneqq 
    \sum_{k=0}^{\infty} 
    \frac{ 
        (a_{1})_{k} \cdots (a_{r})_{k}
    }{
        (b_{1})_{k} \cdots (b_{s})_{k}
    } 
    \frac{z^{k}}{k!} 
    \,\, , 
    \label{eq:hyperg_def}
\end{equation}
where $(c)_{n} \coloneqq \Gamma (c + n) / \Gamma (c)$ is the Pochhammer symbol (shifted factorial). 
In Eq.~(\ref{eq:conj_master}), it can be noticed that $a_{1} = - 2 s$, $a_{2} = \cdots = a_{p+1} = -s$, and $b_{1} = \cdots = b_{p} = 1-s$, so one realises the relation that $a_{1} + 1 = a_{2} + b_{1} = \cdots = a_{p+1} + b_{p} = - 2 s + 1$. 
When such a relation holds, the series is called \textit{well poised}. 
For several $n$'s, Eq.~(\ref{eq:conj_master}) gives 
\begin{subequations}
	\label{eq:conj_exa}
	\begin{align}
		\int \dd y \, y^{2} \cos^{2 s} y 
		&= \frac{e^{- 2 i s y}}{2^{2s} s^{3}} \qty[ 
			i \frac{(s y)^{2}}{2} {}_{2} F_{1} + \frac{(sy)^{1}}{2} {}_{3} F_{2} - i \frac{(sy)^{0}}{4} {}_{4} F_{3} 
		] 
		\,\, , 
        \label{eq:exa2}
		\\ 
		\int \dd y \, y^{3} \cos^{2 s} y 
		&= \frac{e^{- 2 i s y}}{2^{2s} s^{4}} \qty[ 
			i \frac{(s y)^{3}}{2} {}_{2} F_{1} + \frac{4}{3} (sy)^{2} {}_{3} F_{2} - i \frac{3}{4} (sy)^{1} {}_{4} F_{3} - \frac{3}{8} (sy)^{0} {}_{5} F_{4}  
		] 
		\,\, , 
        \label{eq:exa3}
        \\ 
        \int \dd y \, y^{4} \cos^{2 s} y 
		&= \frac{e^{- 2 i s y}}{2^{2s} s^{5}} \qty[ 
            i \frac{(s y)^{4}}{2} {}_{2} F_{1} 
            + (s y)^{3} {}_3 F_{2} 
            - i \frac{3}{2} (s y)^{2} {}_{4} F_{3} 
            - \frac{3}{2} (s y)^{1} {}_{5} F_{4} 
            + i \frac{3}{4} (s y)^{0} {}_{6} F_{5} 
        ] 
        \,\, , 
        \label{eq:exa4}
	\end{align}
\end{subequations}
and so on. 
The arguments of the hypergeometric functions are omitted but follow Eq.~(\ref{eq:conj_master}). 

When the integration (\ref{eq:conj_master}) is carried out from $- \pi/2$ to $\pi/2$, every hypergeometric function is evaluated at $z = 1$. 
In this limit, the infinite summation (\ref{eq:hyperg_def}) reduces to the closed forms in terms of the Gamma functions. 
For instance, there are Gauss'~\cite{gauss1813disquisitiones} and Dixon's~\cite{10.1112/plms/s1-35.1.284} summation formulas for ${}_{2} F_{1}$ and ${}_{3} F_{2}$ respectively given by~\cite{andrews1999special}
\begin{subequations}
    \label{eq:sumf_ex}
    \begin{align}
        {}_{2} F_{1} \qty( ~
            \begin{matrix}
                a, \, b \\ c 
            \end{matrix}
            ~ \middle| ~ 1 
        ) 
        &= \frac{ \Gamma (c) \Gamma (c - a - b) }{ \Gamma (c - a) \Gamma (c - b) } 
        \,\, , 
        \label{eq:sumf_ex1}
        \\ 
        {}_{3} F_{2} \qty( ~
            \begin{matrix}
                a, \, -b, \, -c \\ a + b + 1, \, a + c + 1  
            \end{matrix}
            ~ \middle| ~ 1 
        ) 
        &= \frac{ \Gamma (a/2 + 1) \Gamma (a + b + 1) \Gamma (a + c + 1) \Gamma (a/2 + b + c + 1) }{ \Gamma (a + 1) \Gamma (a/2 + b + 1) \Gamma (a/2 + c + 1) \Gamma (a + b + c + 1) } 
        \,\, . 
        \label{eq:sumf_ex2}
    \end{align}
\end{subequations}

\subsection{$n = 2$}

For $n = 2$, it follows from Eq.~(\ref{eq:exa2}) that 
\begin{equation}
    \int_{- \pi/2}^{\pi/2} \dd y \, y^{2} \cos^{2 s} y 
	= \frac{1}{2^{2s} s^{3}} \qty[
        \frac{ (s \pi)^{2} }{4} \sin (s \pi) {}_{2} F_{1} ( \cdots \mid 1 ) 
        + \frac{s \pi}{2} \cos (s \pi) {}_{3} F_{2} (\cdots \mid 1) 
        - \frac{1}{2} \sin (s \pi) {}_{4} F_{3} ( \cdots \mid 1 )
    ] \,\, . 
    \label{eq:defint_2}
\end{equation}
The arguments of the hypergeometric functions other than $z = 1$ are again omitted for shorthand notations. 
The summation formulas such as Eqs.~(\ref{eq:sumf_ex}) enable us to reduce the above expression through 
\begin{subequations}
    \label{eq:sumf_sumf}
    \begin{align}
        {}_{2} F_{1} ( \cdots \mid 1 ) 
        &= \frac{2^{2 s}}{\sqrt{\pi}} \Gamma ( 1 - s) \Gamma \qty( s + \frac{1}{2} ) 
        \,\, , 
        \\ 
        {}_{3} F_{2} ( \cdots \mid 1 ) 
        &= \frac{2^{2 s}}{\sqrt{\pi}} \Gamma ( 1 - s) \Gamma \qty( s + \frac{1}{2} )  \frac{s \pi}{ \tan ( s \pi) } \,\, , 
        \\ 
        {}_{4} F_{3} ( \cdots \mid 1 ) 
        &= \frac{2^{2 s}}{\sqrt{\pi}} \Gamma ( 1 - s) \Gamma \qty( s + \frac{1}{2} ) \qty{ 
            1 - \frac{1}{2} (s \pi)^{2} + \qty[ \frac{s \pi}{ \sin (s \pi) } ]^{2} - s^{2} \Psi^{(1)} (s) 
        } \,\, , 
    \end{align}
\end{subequations}
where $\Psi (z)$ is the polygamma function and $\Psi^{(n)} (z)$ is its $n$-th derivative, 
\begin{equation}
    \Psi (z) 
    \coloneqq \dv{\ln \Gamma (z)}{z} 
    = \frac{ \dd \Gamma (z) / \dd z }{\Gamma (z)} 
    \,\, , 
    \qquad 
    \Psi^{(n)} (z) 
    \coloneqq \dv[n]{\Psi (z)}{z} 
    = \frac{ \dd^{n+1} \ln \Gamma (z) }{ \dd z^{n+1} } 
    \,\, . 
\end{equation}
Combining Eqs.~(\ref{eq:sumf_sumf}), the integral (\ref{eq:defint_2}) reduces to 
\begin{equation}
    \int_{- \pi/2}^{\pi/2} \dd y \, y^{2} \cos^{2 s} y 
    = \frac{\sqrt{\pi}}{2} \frac{ \Gamma (s + 1/2) }{ \Gamma ( s + 1) } \qty[ \Psi^{(1)} (s) - \frac{1}{s^{2}} ] 
    = \frac{\sqrt{\pi}}{2} \frac{ \Gamma (s + 1/2) }{ \Gamma ( s + 1) } \sum_{k = 1}^{\infty} \frac{1}{(s + k)^{2}} 
    \,\, . 
\end{equation}
One thus finally arrives at the analytical expression for the stationary variance of the field, 
\begin{equation}
    \expval{ \qty( \frac{\phi}{H} )^{2} }_{\infty} 
	= \frac{3}{8 \pi^{2}} \qty[ s \sum_{k=1}^{\infty} \frac{1}{(s + k)^{2}} ] \qty( \frac{H}{m} )^{2} 
	\,\, . 
	\label{eq:demo_sec}
\end{equation}
Note that in Eq.~(\ref{eq:demo_sec})  $\alpha$ has been identified with the mass of the test field by $\alpha = (1 / \sqrt{6 s} \, ) (m / H)$, as was done in the main text. 
Comparing Eqs.~(\ref{eq:demo_sec}) and~(\ref{eq:exs_mv}), the quantity in the square bracket in Eqs.~(\ref{eq:demo_sec}) measures the deviation of the variance from that of the harmonic-oscillator case. 

\begin{figure}
    \centering
    \includegraphics[width = 0.8\linewidth]{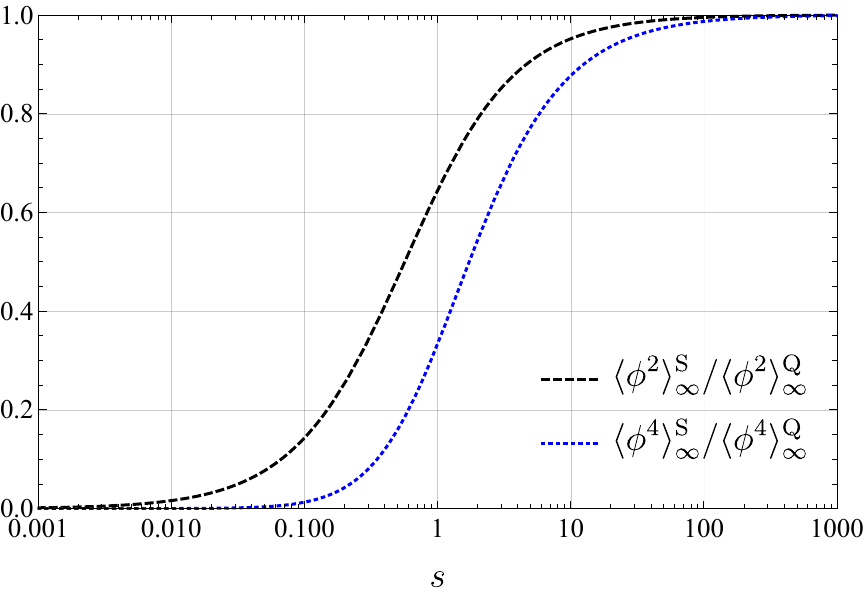}
    \caption{
        The deviation of the second and fourth statistical moments of $\phi$ confined in the trigonometric potential, $v (y) = - s \ln \cos y$, from the case where the curvaton is confined in the pure quadratic potential. 
    }
    \label{fig:apx_1}
\end{figure}

The dashed black curve in Figure~\ref{fig:apx_1} shows this difference for a wide range of $s$. 
It is a monotonically increasing function that approaches to unity as $s \to \infty$, that is, 
\begin{equation}
	\frac{ 
		\displaystyle \expval{ \phi^{2} }_{\infty}^{\text{S}} 
	}{ 
		\displaystyle \expval{ \phi^{2} }_{\infty}^{\text{Q}} 
	} 
	= s \sum_{k=1}^{\infty} \frac{1}{(s + k)^{2}} 
	~~ \xrightarrow{s \to \infty} ~~ 
	1 \,\, . 
\end{equation}
The superscripts ``S'' and ``Q'' indicate the trigonometric Scarf and quadratic potentials, respectively. 
In other words, when $s \to \infty$, the test field cannot tell whether it is confined in the quadratic or the trigonometric Scarf potential. 
This is because in the $s \to \infty$ limit the trigonometric Scarf potential $v (y) = - s \ln \cos y$ behaves as the Dirac-$\delta$ function centred at $y = 0$, prohibiting $\phi$ from going away from the bottom. 
For $0 < s < \infty$, on the other hand, the Scarf potential has a steeper gradient compared to that of the quadratic potential, which explains that the deviation factor in the variance is always less than unity. 

\subsection{$n = 4$}
\label{subsec:n4}

For $n = 4$, it follows from Eq.~(\ref{eq:exa4}) that 
\begin{align}
    \int_{- \pi/2}^{\pi/2} \dd y \, y^{4} \cos^{2 s} y 
	&= \frac{1}{2^{2s} s^{5}} \left[
        \frac{1}{16} (s \pi)^{4} \sin (s \pi) {}_{2} F_{1} ( \cdots \mid 1 ) 
        + \frac{1}{4} (s \pi)^{3} \cos ( s \pi ) {}_{3} F_{2} ( \cdots \mid 1 ) 
    \right. 
    \notag \\ 
    &\quad 
    \left. 
        - \frac{3}{4} (s \pi)^{2} \sin (s \pi) {}_{4} F_{3} ( \cdots \mid 1 ) 
        - \frac{3}{2} (s \pi) \cos (s \pi) {}_{5} F_{4} ( \cdots \mid 1 ) 
        + \frac{3}{2} \sin (s \pi) {}_{6} F_{5} ( \cdots \mid 1 ) 
    \right] 
    \,\, . 
    \label{eq:defint_4}
\end{align}
The right-hand side can be simplified using Eqs.~(\ref{eq:sumf_sumf}) and  
\begin{subequations}
    \begin{align}
        {}_{5} F_{4} ( \cdots \mid 1 ) 
        &= \frac{2^{2s}}{\sqrt{\pi}} 
        \Gamma (1 - s) \Gamma \qty( s + \frac{1}{2} ) 
        \frac{ s \pi}{ \tan (s \pi) } 
        \qty{
            1 - \frac{1}{6} (s \pi)^{2} + \qty[ \frac{s \pi}{\sin (s \pi)} ]^{2} - s^{2} \Psi^{(1)} (s) 
        } 
        \,\, , 
        \\ 
        {}_{6} F_{5} ( \cdots \mid 1 ) 
        &= \frac{2^{2s}}{\sqrt{\pi}} 
        \Gamma (1 - s) \Gamma \qty( s + \frac{1}{2} ) 
        \frac{s^{4} D_{1} (s)}{24} 
        \,\, . 
    \end{align}
\end{subequations}
The function $D_{1} (s)$ is defined by 
\begin{align}
    D_{1} (s) 
    &\coloneqq \qty[ \Psi^{(0)} (-s) ]^4 - 4 \Psi ^{(0)} (s+1) \qty[ \Psi^{(0)}(-s) ]^3 
    + 6 \qty{ 
        \qty[ \Psi^{(0)} (s+1) ]^2 + \Psi^{(1)} (-s) - \Psi^{(1)} (s+1) 
    } \qty[ \Psi^{(0)} (-s) ]^2 
    \notag \\ 
    &\quad - 4 \qty{ 
        \qty[ \Psi^{(0)} (s+1) ]^3 + 3 \qty[ \Psi^{(1)} (-s) - \Psi ^{(1)} (s+1) ] \Psi^{(0)} (s+1) - \Psi^{(2)} (-s) 
        + \Psi^{(2)} (s+1) 
    } \Psi^{(0)} (-s) 
    \notag \\ 
    &\quad 
    + \qty[ \Psi^{(0)} (s+1) ]^4 
    + 3 \qty[ 
        \Psi^{(1)} (-s) - \Psi^{(1)} (s+1) 
    ]^2 
    + 6 \qty[ \Psi^{(0)} (s+1) ]^2 \qty[ 
        \Psi^{(1)} (-s) - \Psi^{(1)} (s+1)
    ] \notag \\ 
    &\quad + \Psi^{(3)} (-s) - \Psi^{(3)} (s+1) - \frac{8 \pi^{3}}{\tan (s \pi) \sin^{2} (s \pi)} \Psi ^{(0)} (s+1) 
    \,\, . 
\end{align}
Combining them all, one arrives at 
\begin{equation}
	\expval{ \qty( \frac{\phi}{H} )^{4} }_{\infty} 
	= \frac{27}{64 \pi^{4}} \qty[ \frac{ D_{2} (s) }{12 s^{2}} ] \qty( \frac{H}{m} )^{4} 
	\,\, . 
	\label{eq:demo_four}
\end{equation}
The function $D_{2} (s)$ is defined by 
\begin{align}
    D_{2} (s) 
    &\coloneqq (s \pi)^{4} + 4 (s \pi)^{2} \qty[ \frac{s \pi}{\tan (s \pi)} ]^{2} - 6 (s \pi)^{2} \qty{ 
        2 - (s \pi)^{2} + 2 \qty[ \frac{s \pi}{\sin (s \pi)} ]^{2} - 2 s^{2} \Psi' (s) 
    } 
    \notag \\ 
    &\quad - 4 \qty[ \frac{ s \pi }{ \tan (s \pi) } ]^{2} \qty{
        6 - (s \pi)^{2} + 6 \qty[ \frac{s \pi}{\sin (s \pi)} ]^{2} - 6 s^{2} \Psi' (s) 
    } 
    + s^{4} D_{1} (s) 
    \,\, . 
\end{align}
Comparing Eq.~(\ref{eq:demo_four}) with the Gaussian case where $\expval{ \phi^{4} } = 3 \expval{ \phi^{2} }^{2}$, the factor $D_{2} (s) / 12 s^{2}$ measures the deviation from the pure harmonic oscillator case. 
The dotted blue curve in Figure~\ref{fig:apx_1} shows it, 
\begin{equation}
    \frac{ 
		\displaystyle \expval{ \phi^{4} }_{\infty}^{\text{S}} 
	}{ 
		\displaystyle \expval{ \phi^{4} }_{\infty}^{\text{Q}} 
	} 
	= \frac{D_{2} (s)}{12 s^{2}} 
	~~ \xrightarrow{s \to \infty} ~~ 
	1 \,\, , 
\end{equation}
which again goes to unity in the limit $s \to \infty$ as expected. 

\bibliography{Bibliography}
\bibliographystyle{JHEP}

\end{document}